\documentstyle[12pt,axodraw,epsfig]{article}
\def\beq{\begin{equation}}
\def\ra{\rightarrow}
\def\eeq{\end{equation}}

\def\bea{\begin{eqnarray}}
\def\eea{\end{eqnarray}}

\begin{document}

\begin{figure}[t]  
\vspace*{-0.4 cm}
\psfig{figure=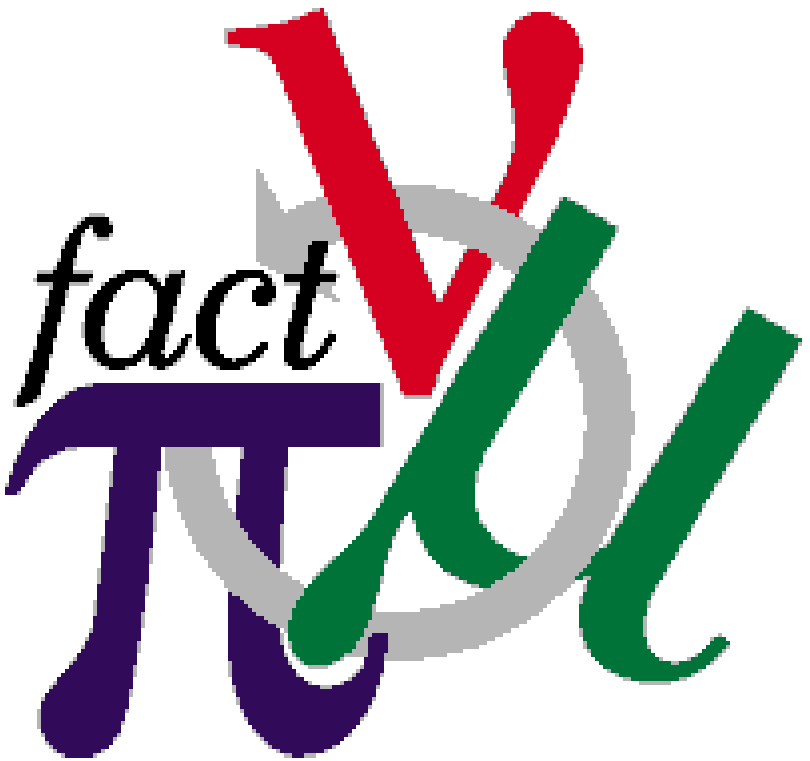,height=1cm,width=1cm,angle=0}
\vspace*{-2.3 cm}
\end{figure}

\title{\begin{flushright}
\vspace*{-1.2 cm}
{\normalsize hep-ph/0008276} \\
\vspace*{-0.4 cm}
{\normalsize CERN-NUFACT note 43}\\
\vspace*{-0.4 cm}
{\normalsize CERN-TH/2000-213, FISIST/8-2000/CFIF}
\\
\end{flushright} 
\vspace*{0.7 cm} {\Large \bf 
Charged-Lepton-Flavour Violation in Kaon Decays in Supersymmetric
Theories}}
\author{{\normalsize 
\bf\hspace{-1.3cm} A. Belyaev$^a$, M. Chizhov$^{b,c}$,
A. Dorokhov$^d$, J. Ellis$^c$, M. E. G\'omez$^e$}
{\normalsize and \bf S. Lola$^c$ }
\vspace*{0.3 cm}
\\
{\small\it\hspace{-1.6cm} a) Skobeltsyn Institute for Nuclear Physics,
Moscow State University,}\\
{\small\it\hspace{-1.6cm} 119 899 Moscow, Russia}\\
{\small\it\hspace{-1.6cm} b) Centre for Space Research and Technologies, 
Faculty of Physics, University of Sofia,}\\
{\small\it\hspace{-1.6cm} 1164~Sofia, Bulgaria}\\
{\small\it\hspace{-1.6cm}
c) Theory Division, CERN, CH-1211 Geneva 23, Switzerland } \\ 
{\small\it\hspace{-1.6cm} \it d) Bogoliubov Laboratory for Theoretical
Physics, Joint Institute for Nuclear Research,} \\
{\small\it\hspace{-1.6cm} 141 980 Dubna, Russia}\\
{\small\it\hspace{-1.6cm} e) CFIF, Departamento de Fisica,
Instituto Superior T\'ecnico,
Av. Rovisco Pais,} \\
{\small\it\hspace{-1.6cm} 1049-001~Lisboa, Portugal} \\
}
\date{}
\maketitle
\vspace*{-0.5 cm}
{\bf Abstract:
{\small 
We discuss rare kaon decays that violate charged-lepton flavour
conservation in supersymmetric theories with and without $R$~parity,
in view of possible experiments using an intense proton source as
envisaged for a neutrino factory. In
the Minimal Supersymmetric Standard Model, such decays are generated by
box
diagrams involving charginos and neutralinos, but the limits from $\mu
\rightarrow e \gamma$, $\mu$--$e$ conversion and $\Delta m_K$ constrain
the branching ratios to challengingly small values.
However, this is no longer the
case in $R$-violating theories, where such decays may occur at tree level
at rates close to the present experimental limits.
Within this framework, we obtain bounds on products of $LL\bar{E}$ and
$LQ\bar{D}$ operators from the experimental upper limits on $K^0 
\rightarrow
\mu^\pm e^\mp$ and $K^{\pm,0} \rightarrow \pi^{\pm,0} \mu^\pm e^\mp$
decays. We also note the possibility
of like-sign lepton decays $K^\pm \rightarrow \pi^\mp \ell^\pm \ell^\pm$
in the presence of non-zero $\tilde{b}_L$--$\tilde{b}_R$ mixing. We
conclude that rare kaon decays violating charged-lepton flavour
conservation could be an interesting signature of $R$ violation. 
}}

\begin{center}
{\it Prepared for the Kaon Physics Working Group 
as part of the \\
ECFA/CERN study of \\
NEUTRINO FACTORY \& MUON STORAGE RINGS AT CERN}
\end{center}

\section{Introduction}

The recent Super-Kamiokande data~\cite{SuKa} have
triggered an upsurge of interest in extensions of the
Standard Model with massive neutrinos
and/or violation of the charged-lepton numbers
in processes such as 
$\mu \rightarrow e \gamma$,
$\mu \rightarrow 3 e$,
$\tau \rightarrow \mu \gamma$ and $\mu \to e$
conversion on heavy nuclei~\cite{neutrinoLFVns,neutrinoLFVs,a,GELLN,KO}.
The present experimental upper bounds on the most
interesting of these processes are:
\begin{eqnarray}
BR(\mu \ra e \gamma) \, < \, 1.2 \times 10^{-11} \, &:& \, \,
\cite{Brooks} \nonumber\\
BR(\mu^+ \ra e^+ e^+ e^-) \, < \, 1.0 \times 10^{-12} \, &:& \, \,
\cite{Bellgardt} \nonumber\\
R(\mu^- Ti \ra e^- Ti) \, < \, 6.1 \times 10^{-13} \, &:& \, \,
\cite{Wintz} \nonumber\\
BR(\tau \ra \mu \gamma) \, < 1.1 \times \, 10^{-6} \, &:& \, \,
\cite{CLEO}
\label{processes}
\end{eqnarray}
Our main interest in this paper is in rare kaon decays, for which the
current experimental bounds are:
\begin{eqnarray}
BR(K^0_L \ra \mu^\pm e^\mp) \, < 4.7 \times \, 10^{-12} \, &:& \, \,
\cite{kaon1}\nonumber \\
BR(K^0_L \ra e^+ e^-) \, = (8.7^{+5.7}_{-4.1})
\times \, 10^{-12} \, &:& \, \,
\cite{kaon1'}\nonumber \\
BR(K^0_L\ra \mu^+ \mu^-) \, = (7.18\pm0.17) 
\times \, 10^{-9} \, &:& \, \,
\cite{kaon1''} \nonumber\\
BR(K^+ \ra \pi^+ \mu^+ e^-) \, < 2.8 \times \, 10^{-11} \, &:& \, \,
\cite{kaon2}\nonumber\\
BR(K^+ \ra \pi^+ \mu^- e^+) \, 
< 5.2 \times \, 10^{-10} \, &:& \, \,
\cite{kaon4}\nonumber\\
BR(K^0_L \ra \pi^0 \mu^\pm e^\mp) \,
< 3.1 \times \, 10^{-9} \, &:& \, \,
\cite{kaon02''}\nonumber\\
BR(K^+ \ra \pi^+ e^+ e^-) \, < (2.94\pm0.05\pm0.14) 
\times \, 10^{-7} \, &:& \, \,
\cite{kaon2''}\nonumber\\
BR(K^0_L \ra \pi^0 e^+ e^-) \, < 4.3 \times \, 10^{-9} \, &:& \, \,
\cite{kaon02}\nonumber\\
BR(K^+ \ra \pi^+ \mu^+ \mu^-) \, < (7.6\pm2.1) 
\times \, 10^{-8} \, &:& \, \,
\cite{PDG}\nonumber\\
BR(K^0_L \ra \pi^0 \mu^+ \mu^-) \, < 5.1 \times \, 10^{-9} \, &:& \, \,
\cite{kaon02'}\nonumber\\
BR(K^+ \ra \pi^- \mu^+ e^+) \, 
< 5.0 \times \, 10^{-10} \, &:& \, \,
\cite{kaon4}\nonumber\\
BR(K^+ \ra \pi^- e^+ e^+) \,     < 6.4 \times \, 10^{-10} \, &:& \, \,
\cite{kaon4}\nonumber\\
BR(K^+ \ra \pi^- \mu^+ \mu^+) \, < 3.0 \times \, 10^{-9} \, &:& \, \,
\cite{kaon4}
\label{Kprocesses}
\end{eqnarray}
Projects are being proposed that could be used to improve these limits
significantly, e.g.,
some of the powerful proton
sources being proposed for neutrino factories~\cite{mufact} could
provide intense secondary kaon beams.

Any observable rate for one of these processes would constitute
unambiguous evidence for new physics.
The rates for such processes remain extremely suppressed if we simply
extend the Standard Model to include right-handed neutrinos, but
larger rates are possible in more ambitious extensions of the
Standard Model.
Supersymmetry is one example of new physics that could amplify
rates for some of the rare processes (\ref{processes}, \ref{Kprocesses}),
either
in the minimal supersymmetric extension of the Standard Model (MSSM)
or in its modification to include violation of $R$ parity.

In previous works, rare charged-lepton decays and anomalous $\mu \to e$
conversions on heavy nuclei have received considerable
attention~\cite{KO}, leading
to a good understanding of the correlations between the predicted rates
and possible violations of leptonic universality and/or exotic Yukawa
couplings with $\Delta L \neq 0$. It is natural to include rare kaon
decays in this analysis, because strangeness-changing decays occur in the
Standard Model. As mentioned above, there is now considerable discussion
of intense
proton sources to be used as muon sources for neutrino
factories~\cite{mufact}. If the
protons have high energy above $\simeq 15$~GeV, as in some
neutrino factory designs, they would also be copious sources of kaons.
These might provide a new opportunity to study rare $K$ decays with high
statistics, and we are interested to know whether these might cast
additional light on neutrino masses and mixing. 
Specifically, as we show in this paper, rare $K$ decays
that violate charged-lepton number allow one, in the context of the MSSM,
to correlate the (s)quark and (s)lepton mixing, whilst in $R$-violating
supersymmetry one can probe interesting products of Yukawa couplings.

This paper is organized as follows. In Section 2, we analyze the decays
$K^0 \rightarrow \mu^\pm e^\mp$ 
in the MSSM,
finding rates that are relatively small,  although
for certain model parameters they may be within the 
reach of an imaginable kaon beam at a neutrino factory. In Section 3, we
analyze the same
decays in $R$-violating supersymmetric models, finding that the rates
could in principle be much larger, close to the present experimental
limits. Indeed, these existing upper
limits on these decays already impose interesting upper limits on products
of
$LL\bar{E}$ and $LQ\bar{D}$ couplings. We also point out
that $\tilde{b}_L$--$\tilde{b}_R$ mixing in the presence of $R$ violation
could lead to the like-sign-lepton decays $K^\pm \rightarrow \pi^\mp
\ell^\pm \ell^\pm$, although the existing experimental limits on these
processes do not impose interesting upper limits on couplings. Taken
together, our results indicate that rare
$K$ decays could provide interesting signatures for supersymmetric models,
in particular those with $R$ violation. 

\section{Rare Kaon Decays in the MSSM}

In this Section, we evaluate the
rates for rare kaon decays in the MSSM with massive neutrinos,
using the see-saw mechanism~\cite{seesaw}, which we consider to be the
most natural way to obtain neutrino masses
in the sub-eV range. In particular, we assume
Dirac neutrino masses $m_{\nu}^D$ of the same order as
the charged-lepton and quark masses, and heavy Majorana
masses $M_{\nu_R}$, leading to a light effective neutrino mass
matrix of the form:
\begin{equation}
m_{eff}
=m^D_{\nu}\cdot (M_{\nu_R})^{-1}\cdot m^{D^{\normalsize T}}_{\nu}
\label{eq:meff}
\end{equation}
Neutrino-flavour mixing~\cite{MNS} may then occur through either the
Dirac matrix $m^{D}_{\nu}$ and/or the Majorana mass matrix $M_{\nu_R}$,
which may also
feed flavour violation through to the charged leptons.
In non-supersymmetric models with
massive neutrinos, the amplitudes for
charged-lepton-flavour violation are 
proportional to inverse powers of
the right-handed neutrino mass scale 
$M_{\nu_R}$, and thus
the rates for rare decays are extremely
suppressed~\cite{neutrinoLFVns}.
On the other hand, in supersymmetric models one must
also take the dynamics of sneutrinos $\tilde \nu$ into account,
and these processes may only be only suppressed
by inverse powers of the supersymmetry-breaking scale, which 
characterizes $m_{\tilde \nu}$ and is 
at most ${\cal O} (1)$~TeV \cite{neutrinoLFVs}.

The magnitudes of the predicted rates depend on the
details of the masses and mixings of sparticles including the
sneutrinos. If their soft supersymmetry-breaking masses
are non-universal at $M_{GUT}$,
large rates are in general predicted~\cite{susyFC}.
However, even if the soft supersymmetry-breaking masses
are universal at ${M_{GUT}}$,
renormalization effects in the
MSSM with right-handed neutrinos
spoil this diagonal form ~\cite{neutrinoLFVs,a,GELLN} at
lower scales. This is because
the Dirac neutrino and charged-lepton Yukawa
couplings and masses $m_{\ell}$ cannot, in general, be diagonalized
simultaneously. Since both these sets of lepton Yukawa
couplings appear in the renormalization-group equations, 
the lepton and slepton mass matrices also may not be 
diagonalized simultaneously at low energies.

To illustrate this point, let us consider
the simplest example of a 
model based on Abelian flavour symmetries and
symmetric mass matrices \cite{IR}.
Requiring large
(2-3) mixing in this model \cite{GGR}
severely constrains the possible flavour charges
and thus the forms of the charged-lepton and the
neutrino mass matrices.
A representative example is given by Ansatz A of~\cite{GELLN}:
\bea
m_{\ell }  \propto  \left( 
\begin{array}{ccc}
\bar{\epsilon}^{7} & \bar{\epsilon}^{3} & \bar{\epsilon}^{7/2} \\ 
\bar{\epsilon}^{3} & \bar{\epsilon} & \bar{\epsilon}^{1/2} \\ 
\bar{\epsilon}^{7/2} & \bar{\epsilon}^{1/2} & 1
\end{array}
\right),
m^D_{\nu} \propto \left( 
\begin{array}{ccc}
{\epsilon}^{7} & {\epsilon}^{3} & {\epsilon}^{7/2} \\ 
{\epsilon}^{3} & {\epsilon} & {\epsilon}^{1/2} \\ 
{\epsilon}^{7/2} & {\epsilon}^{1/2} & 1
\end{array}
\right)
\label{Amasses}
\eea
where $\bar{\epsilon} = \sqrt{\epsilon} = 0.2$.
We already see that the two matrices
cannot be simultaneously diagonalized and indeed,
\bea
V_\ell = \left(
\begin{array}{ccc}
1 & \bar{\epsilon}^{2} & -\bar{\epsilon}^{7/2} \\
-\bar{\epsilon}^{2} & 1 & \bar{\epsilon}^{1/2} \\
\bar{\epsilon}^{7/2} & -\bar{\epsilon}^{1/2} & 1
\end{array}
\right), V_{\nu_D} = \left(
\begin{array}{ccc}
1 & \bar{\epsilon}^{4} & -\bar{\epsilon}^{7} \\
-\bar{\epsilon}^{4} & 1 & \bar{\epsilon} \\
\bar{\epsilon}^{7} & -\bar{\epsilon} & 1
\end{array}
\right) \label{Asolutions}
\eea
Within this general framework, there is ambiguity in the
specification of numerical coefficients in the matrix elements, which
are expected to be of order unity. We return later to this point.

For a generic texture where the charged lepton and
neutrino matrices are not simultaneously diagonal,
 the slepton mass matrix acquires 
non-diagonal contributions from renormalization at scales below
$M_{GUT}$. In the basis where $m_{\ell}$
is diagonal, these corrections take 
the form \cite{neutrinoLFVs}:
\bea
\delta{m}_{\tilde{\ell}}^2\propto \frac 1{16\pi^2} (3 + a^2)
\ln\frac{M_{GUT}}{M_N}\lambda_D^{\dagger} \lambda_D m_{3/2}^2,
\label{offdiagonal}
\eea
where $\lambda_D$ is the Dirac neutrino Yukawa coupling,
$M_N$ is the scale where the effective neutrino-mass operator is
formed, $a$ is related to
the trilinear mass parameter
$A_\ell \equiv a m_{3/2}$, and
$m_{3/2}$ is the presumed common value $m_0$ of the scalar masses at the
GUT scale. 

In the case of non-universal soft masses, these corrections
are generically negligible. However, the rates for
$\Delta L \ne 0$ processes are generally too large in such
non-universal models~\cite{GELLN}. On the other hand,
models with scalar-mass universality at $M_{GUT}$,
such as no-scale~\cite{Ellis:1984bm}
and gauge-mediated models~\cite{gaugmed}, may yield
acceptable rates for $\Delta L \ne 0$ processes. In such models,
the contributions (\ref{offdiagonal}) related to neutrino masses
dominate and lead to non-negligible rates for the
lepton-flavour-violating processes which are determined by the
off-diagonal terms in the Yukawa matrix $\lambda_N$.
The various different solutions
of the solar neutrino deficit~\cite{bimaximal,abel},
with a small/large mixing angle and with eV or 
$\approx$ 0.03 eV neutrinos, predict in general 
different rates for charged-lepton-flavour violation:
the larger the $\mu-e$ mixing,
and the larger the neutrino mass scales that are required,
the larger the rates. Thus,
degenerate neutrinos with bimaximal mixing may be
expected to yield significantly larger effects than,
for instance, hierarchical neutrinos with
a small vacuum mixing angle. Note, in particular, that
the just-so solutions to the solar neutrino
problem with $\delta m^2 \approx 10^{-10}$ eV$^2$
predict, in the case of hierarchical
neutrino masses, rates that are small,
even if the first/second-generation neutrino mixing is large.

In this MSSM framework, rare kaon decays are generated by
box diagrams involving chargino
and neutralino exchanges. For instance,
for $K^0 \rightarrow \mu^\pm  e^\mp$ we have the
diagrams~\footnote{The contribution 
from the first diagram when the
neutralino is a purely photino state has been
discussed in \cite{In}.} of Fig.~1~\footnote{Kaon 
decays in left-right symmetric
models have been analysed in \cite{Apos}.
In principle, 
using the mixing in the quark sector
- in particular between $s$ and $d$ quarks -
we can also
generate $K \rightarrow \mu e$ by penguin diagrams.
In the cases that all the quark mixing is in the
down sector, or the right-handed mixing in the
down sector is much bigger than the one
in the left (which is the one bounded
by $V_{CKM}$) the rates might be of relevance. However,
we do not address here this
model-dependent issue.}. It is clear that
$K^{\pm,0} \rightarrow \pi^{\pm,0} \mu^\pm  e^\mp$ 
decays can be generated in a similar way, but
in this case the experimental bounds are worse by
almost two orders of magnitude, and we do not discuss them in detail.

\begin{figure*}[h]
\hspace*{0.3 cm}
{\unitlength=1.5 pt
\SetScale{1.5}
\SetWidth{0.7}      
{} \qquad\allowbreak
\begin{picture}(100,80)(0,0)
\ArrowLine(30,20)(0,20)
\Line(30,20)(70,20)
\ArrowLine(100,20)(70,20)
\DashLine(30,20)(30,60){1.0}
\DashLine(70,20)(70,60){1.0}
\ArrowLine(0,60)(30,60)
\Line(30,60)(70,60)
\ArrowLine(70,60)(100,60)
\Text(15.0,70.0)[r]{$s$}
\Text(55.0,70.0)[r]{$\chi^0$}
\Text(90.0,70.0)[r]{$\mu$}
\Text(15.0,10.0)[r]{$d$}
\Text(55.0,10.0)[r]{$\chi^0$}
\Text(90.0,10.0)[r]{$e$}
\Text(25.0,40.0)[r]{$\tilde{d}_i$}
\Text(80.0,40.0)[r]{$\tilde{\ell}_i$}
\end{picture} 
{} \qquad\allowbreak
\begin{picture}(100,80)(0,0)
\ArrowLine(30,20)(0,20)
\Line(30,20)(70,20)
\ArrowLine(100,20)(70,20)
\DashLine(30,20)(30,60){1.0}
\DashLine(70,20)(70,60){1.0}
\ArrowLine(0,60)(30,60)
\Line(30,60)(70,60)
\ArrowLine(70,60)(100,60)
\Text(15.0,70.0)[r]{$s$}
\Text(55.0,70.0)[r]{$\chi^\pm$}
\Text(90.0,70.0)[r]{$\mu$}
\Text(15.0,10.0)[r]{$d$}
\Text(55.0,10.0)[r]{$\chi^\pm$}
\Text(90.0,10.0)[r]{$e$}
\Text(25.0,40.0)[r]{$\tilde{u}_i$}
\Text(80.0,40.0)[r]{$\tilde{\nu}_i$}
\end{picture}
}
\caption{\label{fig1}
\em MSSM box diagrams for $K^0 \rightarrow \mu^\pm e^\mp$.
There is another neutralino exchange
diagram corresponding to the permutation of the $\mu$ and $e$.
Since $\chi^0$ is a Majorana spinor,
there are contributions from the neutralinos that
differ in the number of mass insertions.}
\end{figure*}
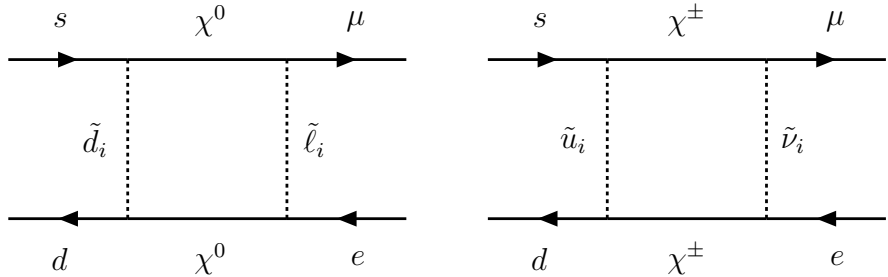

Our procedure for evaluating these contributions is as follows:

\begin{itemize}

\item 
We first find the maximal squark mixing that is allowed
by the neutral-kaon mass difference $\Delta m_K$
\cite{LFVK}.

\item
We next find the maximal slepton mixing allowed by
$\mu \rightarrow e \gamma$ and $\mu$--$e$
conversion in nuclei {\em in a model-independent
way}.

\item
Having fixed these values, finally we calculate the rates
for rare kaon decays.

\end{itemize}

As we noted previously, the $\mu-e$ mixing is constrained
by the form of the neutrino textures and thus by the
recent neutrino data: 
in general, degenerate neutrinos with large angle MSW
oscillations require smaller soft masses to be consistent
with the observed rates. However, as we found in
\cite{GELLN}, even for the small-angle MSW solutions
of the solar neutrino deficit, we can
obtain large rates for values of $m_{0}$ and $m_{1/2}$ well
below 500 GeV. The latest Super-Kamiokande data~\cite{SKnew}
on solar neutrinos favour large mixing angles, which might suggest   
larger $\mu - e$ flavour violation. Thus, considering models with
small mixing angles is conservative.

\begin{figure}
\begin{minipage}[b]{8in}
\epsfig{file=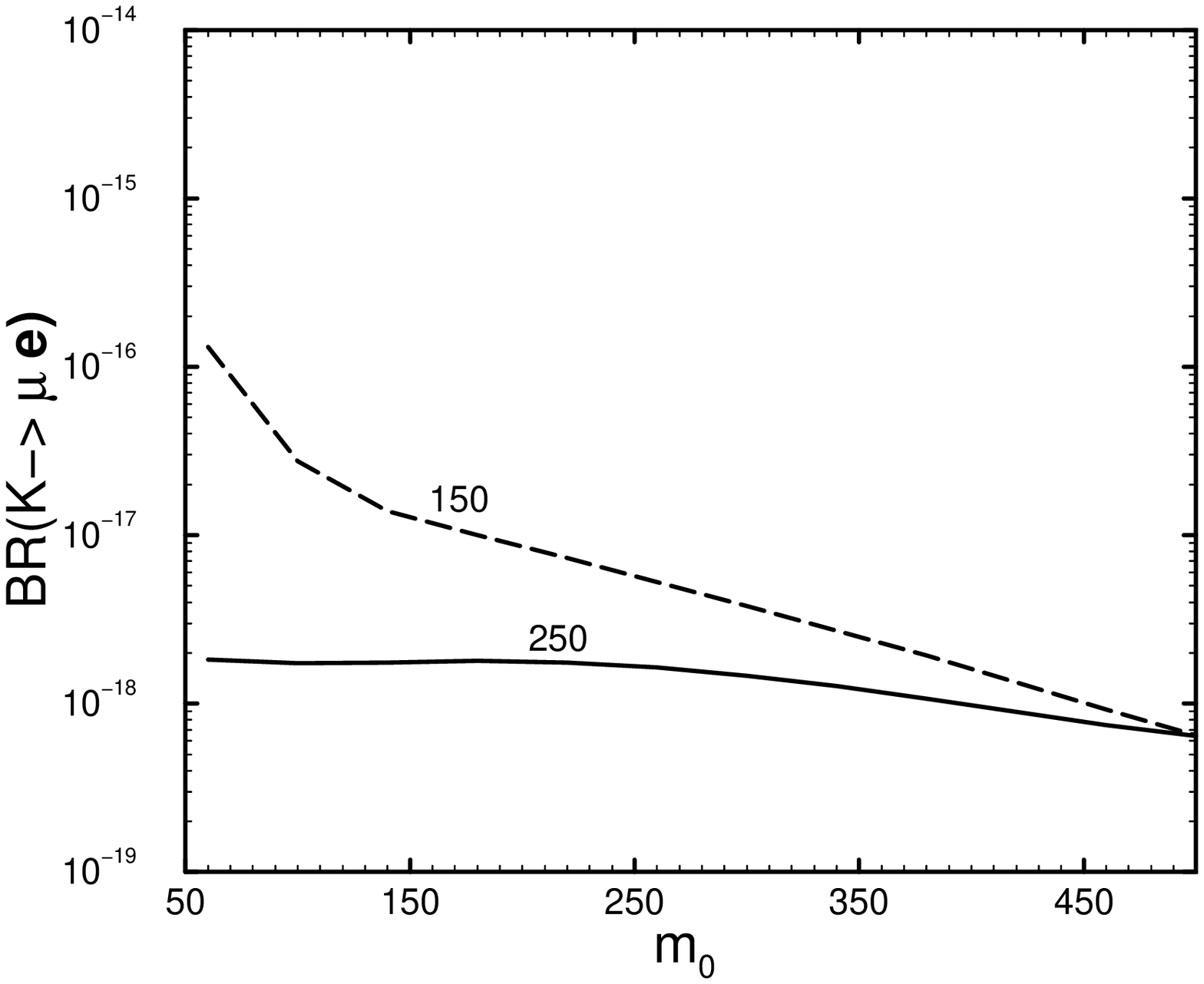,width=3in}
\epsfig{file=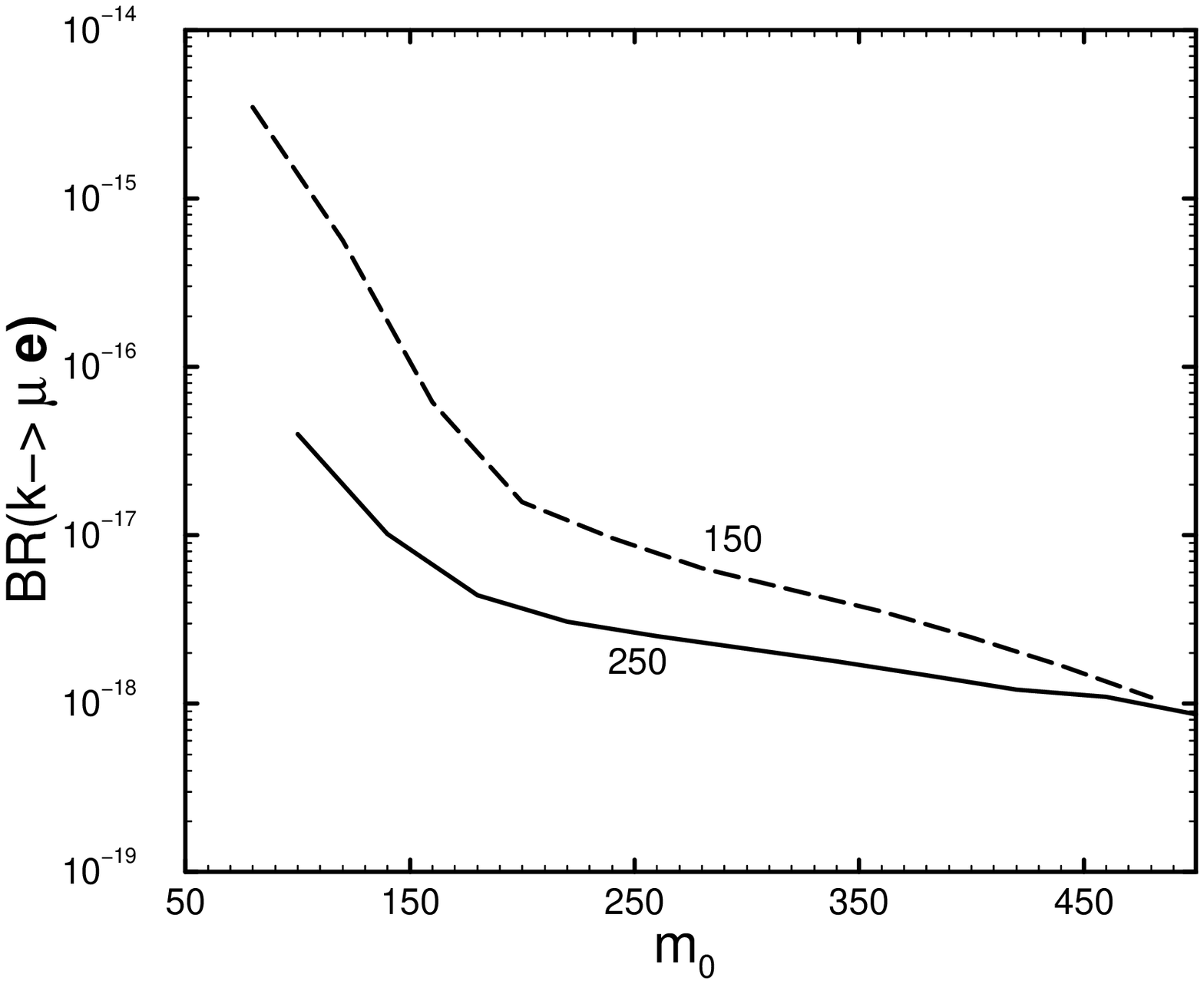,width=3in} \hfill
\end{minipage}
\begin{minipage}[b]{8in}
\epsfig{file=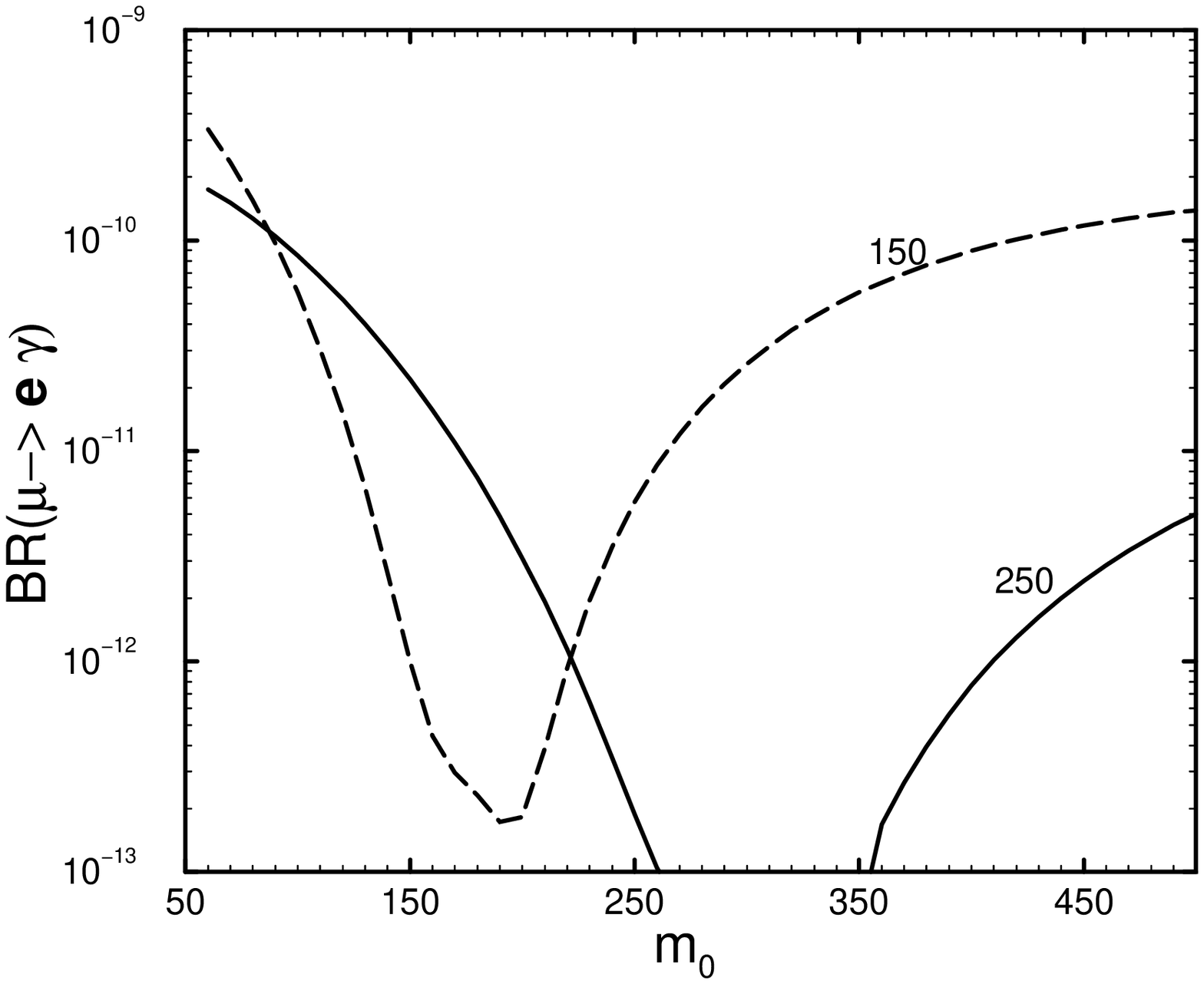,width=3in}
\epsfig{file=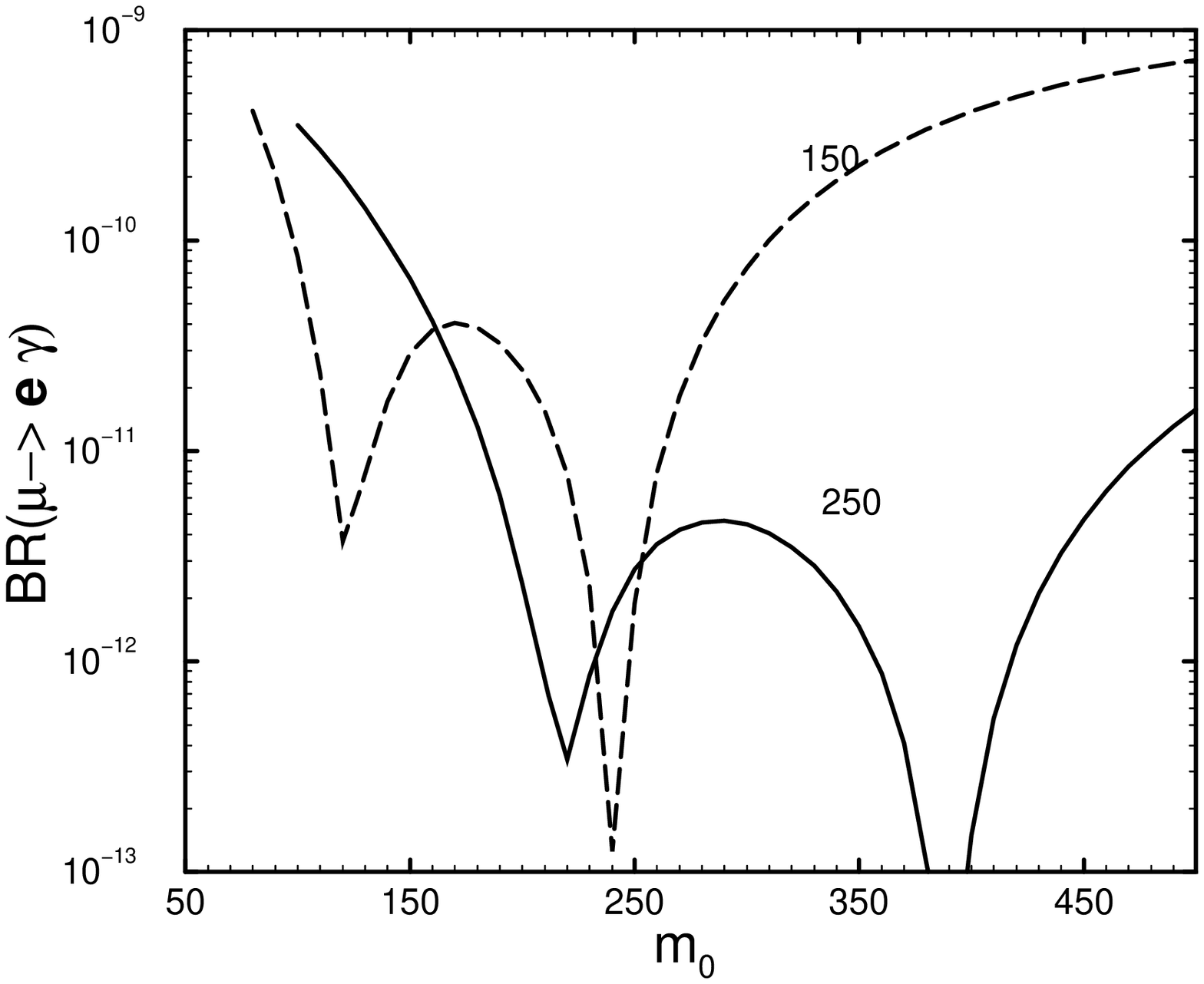,width=3in} \hfill
\end{minipage}
\begin{minipage}[b]{8in}
\epsfig{file=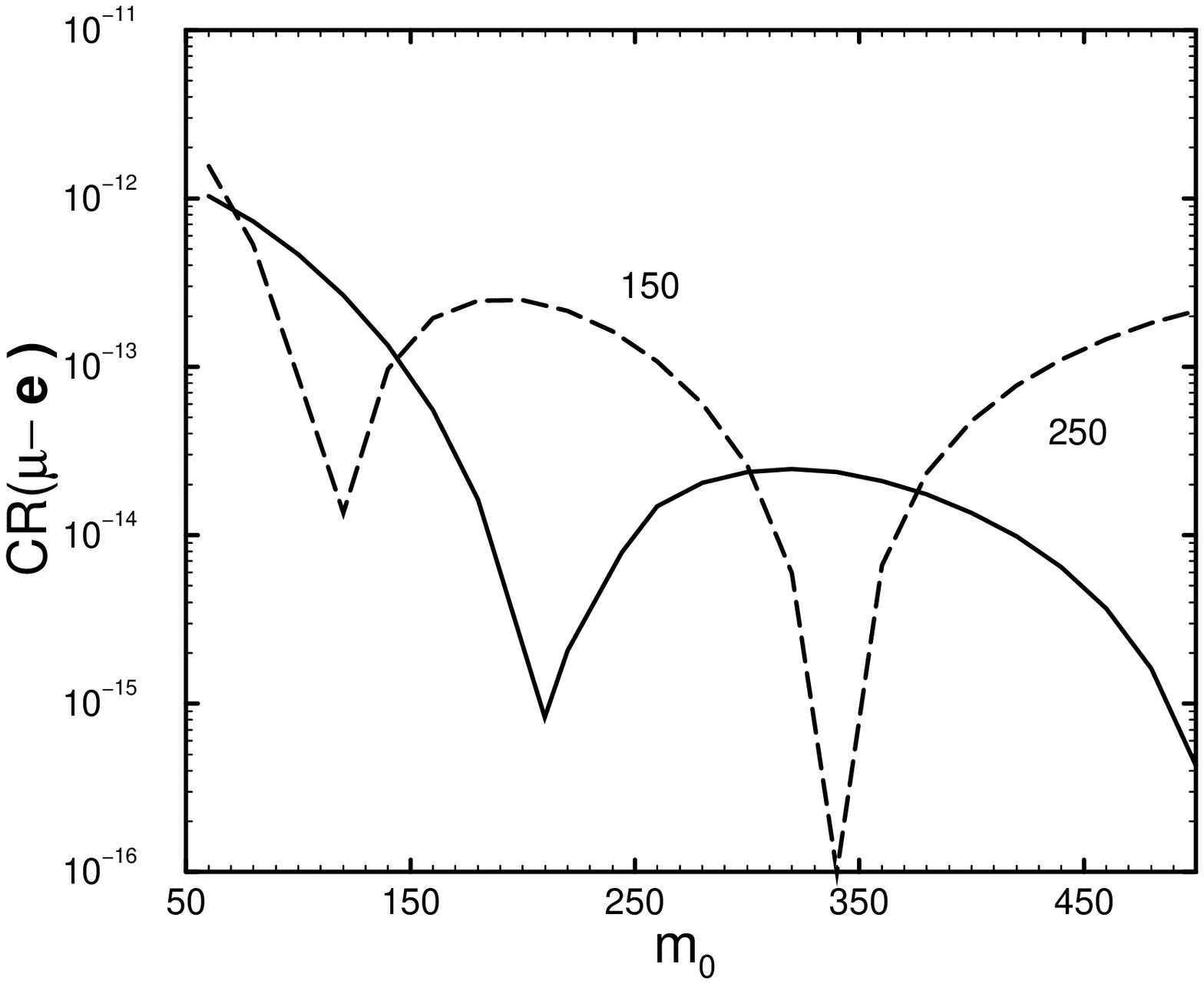,width=3in}
\epsfig{file=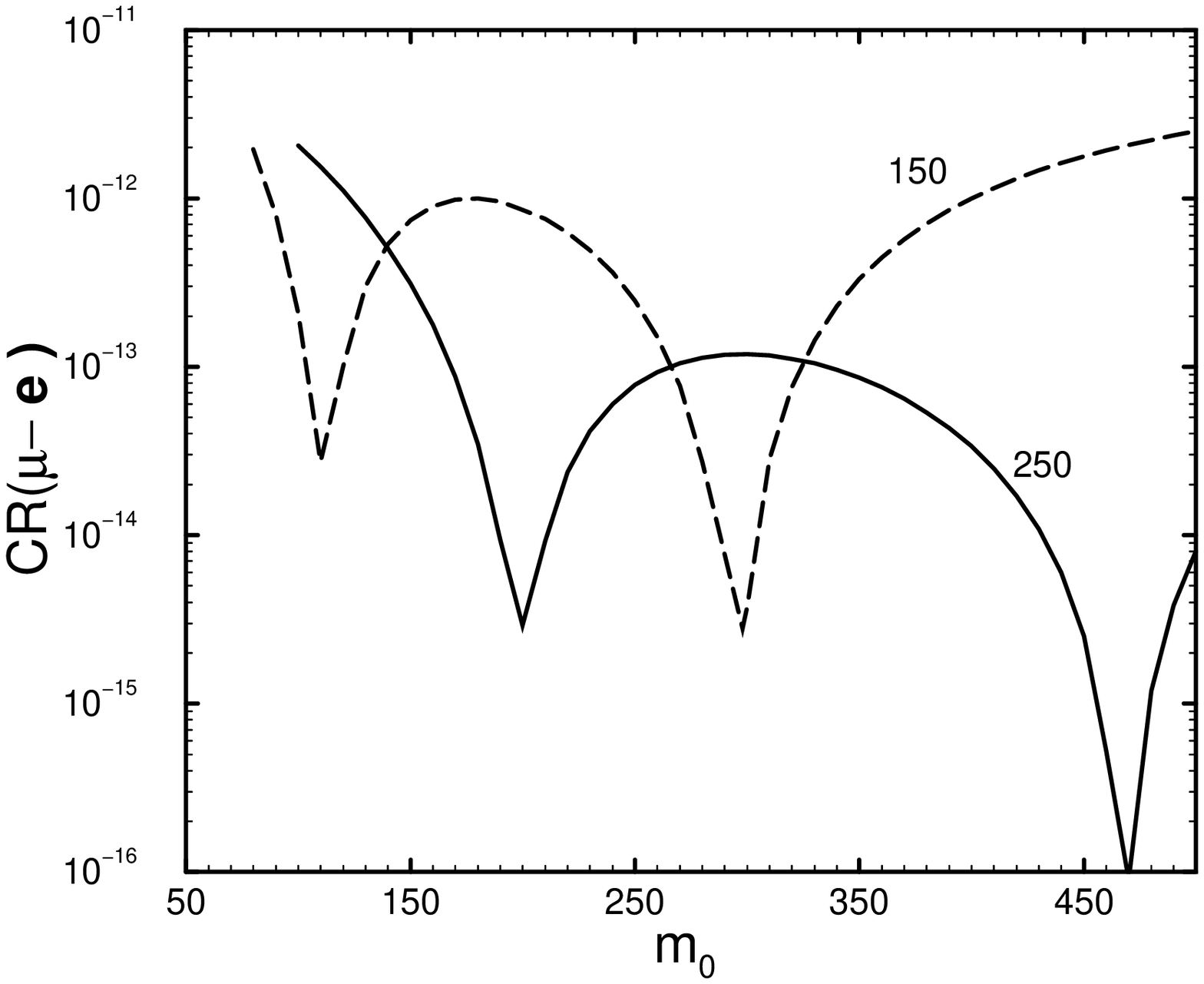,width=3in} \hfill
\end{minipage}
\caption{\it Illustrative predictions for $BR(K \ra \mu e)$,
$BR(\mu \ra  e \gamma)$ and
$BR( \mu-e)$ for different values of $\tan\beta = 10$ (left column), $20$
(right column) and
$m_{1/2}=150$ (dashed lines), $250$ GeV (solid lines), as functions of
$m_0$ (in GeV).}
\end{figure}

For illustration,
we focus here on one such model, namely variant $A_1$ of the
texture (\ref{Amasses}, \ref{Asolutions}), which has the 
numerical values of the ambiguous ${\cal O}(1)$ coefficients fixed as
described in~\cite{GELLN}. For definiteness, we choose its `inverted' option
with negative off-diagonal entries in the Dirac neutrino coupling
matrix~\cite{GELLN}. 
Our results are presented in Figure 2, where we summarise our
predictions for 
$BR(K \ra \mu e)$, in association with the predictions
for 
$BR(\mu \ra  e \gamma)$ and
$BR( \mu-e)$ for different values of $\tan\beta$ and
$m_{1/2}$.
We parameterize the supersymmetric 
masses in terms of the universal GUT-scale
parameters $m_0$  and $m_{1/2}$, for sfermions
and gauginos respectively, and use the renormalization-group
equations of the MSSM to calculate the low-energy sparticle
masses. Other relevant free parameters of
the MSSM are the trilinear coupling $A$
(for which we start with the initial condition
$A_0=- m_{1/2}$),
the sign of the Higgs mixing parameter $\mu$, and
the value of $\tan\beta$. 
Models with different signs of 
$\mu$ give similar results: here we assume $\mu < 0$.

As expected, the larger the value of $\tan\beta$ and
the smaller the soft supersymmetric terms, the larger
the branching ratios, apart from certain cancellations. In the case
$\tan\beta = 10$ and $m_{1/2} = 250$ GeV,
we see that, for the range $m_0 \ge 170$~GeV where
$BR(\mu \ra  e \gamma)$ and
$BR( \mu-e)$ conversion are consistent with the
current experimental bounds (\ref{processes}),
$BR(K \ra \mu e)$ is at most $2 \cdot 10^{-18}$.
However, for the same value of $m_{1/2}$, 
when $\tan\beta = 20$ we find a significantly larger
branching ratio at small values of
$m_0 \sim 170$~GeV. Moreover, for smaller $m_{1/2} = 150$~GeV, we gain
almost two
orders of magnitude when we consider $m_{0}$ in the low-mass
window between 100 and 150 GeV. We recall~\cite{GELLN} that
these lower 
values of $m_{1/2}, m_0$ are consistent with accelerator
constraints and generically yield cold dark matter densities
in the range preferred by cosmology~\cite{EFGO}.

We do not discuss here other model textures for the mass matrices.
Rather, our point here has been to demonstrate
that, despite the limits from $\mu \rightarrow e \gamma$,
$\mu$--$e$ conversion 
and $\Delta m_K$, the branching ratio
of $K \rightarrow \mu e$ 
may be within the reach
of the next generation of experiments, 
namely in the range $10^{-16} \rightarrow 10^{-18}$, at least
if $\tan\beta$ is large and
the soft supersymmetry-breaking terms are small.
The sensitivities (\ref{processes}) to $\mu \rightarrow e \gamma$ and
$\mu$--$e$ conversion could be
improved significantly even before a neutrino factory comes into
operation, and such a machine would offer enhanced prospects for
probing them. It is therefore likely
that the best prospects for discovering
charged-lepton flavour violation may be offered by
$\mu-e$ conversion and $\mu-e \gamma$.
However, rare kaon decays provide a complementary
probe which also gives information on the squark
mixing, in the context of the MSSM.

\section{Rare Kaon Decays in $R$-Violating Supersymmetry}

We now discuss kaon decays violating 
charged-lepton flavour beyond the context of the MSSM.
As is well known, the gauge symmetries of the 
MSSM allow additional dimension-four
Yukawa couplings, of the form
\[
\lambda L_{i}L_{j}{\bar{E}}_{k}, \; \;
\lambda ^{\prime }L_{i}Q_{j}{\bar{D}_{k}}, \; \;
\lambda ^{\prime \prime }{\bar{U}_{i}}{\bar{D}_{j}}{\bar{D}_{k}} 
\]
where the $L(Q)$ are the left-handed lepton (quark) superfields, and the
${\bar{E}}$,(${\bar{D}},{\bar{U}}$) are the corresponding right-handed fields. 
If all these couplings were 
present simultaneously in the low-energy Lagrangian, they would generate
unacceptably fast proton decay. Therefore, extra symmetries must be
invoked to forbid all ($R$-parity~\cite{fayet}), or subsets (baryon and
lepton parities~\cite{BLP}) of these couplings.
In the latter case, very interesting baryon- and 
lepton-number-violating processes may occur~\cite{Rpar}.

Imposing electroweak $SU(2)$ and colour $SU(3)$ invariance, one finds that
there are just 45 $R$-violating couplings in total.
Besides proton decay, there are many experimental
constraints on these couplings, both individually and 
in various combinations, from the non-observation of 
modifications to Standard Model processes and of possible
exotic processes~\cite{constraints,RviolK}.
In order to understand  the possible hierarchies of
$R$-violating couplings, models of flavour symmetries have been 
invoked~\cite{MODELS}. 
For instance, it was found in previous~work \cite{UR1} that
theories with symmetric fermion mass textures lead to the
expectation that $R$-violating couplings are
small: $\leq 10^{-3}$-$10^{-4}$ for 100 GeV sfermion masses,
whilst, in models with asymmetric fermion mass textures, dominance by
a single coupling may be permitted, without
however excluding several products of couplings
from being non-negligible.

In this paper, therefore, we allow the general possibility
that several $R$-violating operators may be large, and discuss
the limits on their combinations that are obtainable from kaon decays.
In this class of models, whilst $\mu \rightarrow e \gamma$
occurs at the one-loop level, $\mu \rightarrow 3 e$,
$\mu-e$ conversion, $K^0 \rightarrow \mu^\pm e^\mp$ and 
$K^{\pm,0} \rightarrow \pi^{\pm,0} \mu^\pm e^\mp$ may occur at
tree level via different combinations of couplings.
For instance, in the case of $LQ\bar{D}$ couplings,
$\mu \rightarrow 3 e$ gives the limit
\cite{constraints}
\bea
(L_1 Q_i \bar{D}_j)(L_2 Q_i \bar{D}_j) \leq 10^{-4} 
\left(\frac{m_{\tilde{f}}}{100 ~{\rm GeV}}\right)^2
\eea
whilst  $\mu$--$e$ conversion in Titanium gives: 
\bea
(L_2 Q_1 \bar{D}_k)(L_1 Q_1 \bar{D}_k) \leq 10^{-8}  
\left(\frac{m_{\tilde{f}}}{100 ~{\rm GeV}}\right)^2
\\
(L_2 Q_j \bar{D}_1)(L_1 Q_j \bar{D}_1) \leq 10^{-8} 
\left(\frac{m_{\tilde{f}}}{100 ~{\rm GeV}}\right)^2
\eea
and the bound
\bea
(L_i Q_1 \bar{D}_2)(L_i Q_2 \bar{D}_1) \leq 10^{-9} 
\left(\frac{m_{\tilde{f}}}{100 ~{\rm GeV}}\right)^2
\eea
is obtainable from $\Delta m_K$~\cite{RviolK}.

What is the connection of these results with neutrino masses?
If $R$ parity is violated, neutrino masses are generated via
one loop diagrams involving the vertices
$ \nu_i \bar{d}_k 
\tilde{d}_{j}, \bar{\nu}^c_i d_j \tilde{d}^*_{k}$ for
$LQ\bar{D}$ operators and the
vertices
$ \nu_i \bar{e}_k 
\tilde{\ell}_{j}, \bar{\nu}^c_i \ell_j \tilde{e}^*_{k}$ for
$LL\bar{E}$ operators. Focusing on $LQ\bar{D}$ couplings
and assuming that the left-right squark soft mixing terms 
are diagonal in the physical basis and proportional
to the associated quark mass,
the induced masses are given by
\begin{equation} 
m_{\nu_{ii'}} \simeq {{n_c \lambda'_{ijk} \lambda'_{i'kj}}
\over{16\pi^2}} m_{d_j} m_{d_k}
\left ( \frac{f(m^2_{d_j}/m^2_{\tilde{d}_k})} {m_{\tilde{d}_k}} +
\frac{f(m^2_{d_k}/m^2_{\tilde{d}_j})} {m_{\tilde{d}_j}}\right ),
\label{mass}
\end{equation}     
where $f(x) = (x\ln x-x+1)/(x-1)^2$, $m_{d_i}$ is the down quark
mass of the $i$th generation inside the loop, $m_{\tilde{d}_i}$ is an
average of $\tilde{d}_{Li}$ and $\tilde{d}_{Ri}$ squark masses, and
$n_c = 3$ is the colour factor. 

Requiring that these contributions be consistent
with the neutrino data gives bounds on the
associated $R$-violating products.
We can see from the above expression that the
heavier the fermions in the loop (including
the associated fermion mass arising from the soft mixing term),
the stricter the bounds \cite{bhat}.
For example, demanding 
$m_{e\mu}<2.5$~eV for sparticle masses
of 100 GeV leads to 
$\lambda^{'}_{133}\lambda^{'}_{233} \leq 3.8 \cdot 10^{-7}$, whilst
for
$\lambda^{'}_{122}\lambda^{'}_{222}$  the bound
drops to  $2.3 \cdot 10^{-4}$ \cite{bhat}. For higher
sfermion masses, larger $R$-violating couplings are allowed.
In Super-Kamiokande-friendly solutions with hierarchical
neutrinos the bounds are stricter by two orders of magnitude.

Note, however, that neutrino masses do not strictly constrain
$K \rightarrow \mu e$ (and in certain cases the rest
of the flavour-violating-processes), since:

$\bullet$
Neutrino masses may only constrain products of
$LL\bar{E}$ or $LQ\bar{D}$ operators,
not mixed $LL\bar{E}$-$LQ\bar{D}$ products.

$\bullet$
Even for the diagrams with products of only $LQ\bar{D}$ operators,
rare kaon decays involve quarks of the lightest
and second-lightest generations. In this case
the bounds from neutrino masses are significantly 
weaker, and the stricter limits come from the
current measurements of the rare kaon decays themselves.
The same is true for $\mu \rightarrow e$ 
conversion and even for $\mu \rightarrow e \gamma$
via $LL\bar{E}$ couplings. For fermions of the
first two generations, the bounds from the lepton-flavour-violating
processes themselves tend to dominate.

\begin{figure}[h]
\hspace*{-1cm}
\begin{picture}(220,100)(0,-10)
\SetWidth{1.}      
\ArrowLine(40,35)(15,35)    \ArrowLine(15,45)(40,45)   \Text(30,55)[]{$K^0$}
\ArrowArcn(110,5 )(80,150,25) \ArrowArcn(110,75)(80,-25,-150) 
\Text(50,46)[]{$d$}
\Text(50,34)[]{$\bar s$}
\Vertex(180,40){3}
\DashLine(180,40)(210,40){5}
\ArrowLine(240,0)(210,40) \Text(245,0)[l]{$\ell^+$}
\ArrowLine(210,40)(240,80)\Text(245,80)[l]{$\ell^-$}
\Vertex(210,40){3}
\Text(190,50)[]{$\tilde\nu$}
\end{picture}

\vspace*{-3.5cm}
\hspace*{7.5cm}
\begin{picture}(220,100)(0,-10)
\SetWidth{1.}      
\ArrowLine(40,35)(15,35)    \ArrowLine(15,45)(40,45)   \Text(30,55)[]{$K^0$}
\ArrowArcn(110,5 )(80,150,55) \ArrowArcn(110,75)(80,-55,-150) 
\Text(50,46)[]{$d$}
\Text(50,34)[]{$\bar s$}
\DashLine(160,15)(160,65){5} 
\Text(165,40)[l]{$\tilde u$}
\Vertex(160,10){3} \Vertex(160,70){3}
\ArrowLine(190,10)(160,10) \Text(195,10)[l]{$\ell^+$}
\ArrowLine(160,70)(190,70) \Text(195,70)[l]{$\ell^-$}
\end{picture}
\caption{\label{fig:kaon0}
\it Quark/sfermion diagrams 
involving $R$-violating couplings that
yield two-body 
$K^0 \rightarrow \ell^\pm \ell^\mp$ decays.}
\end{figure}
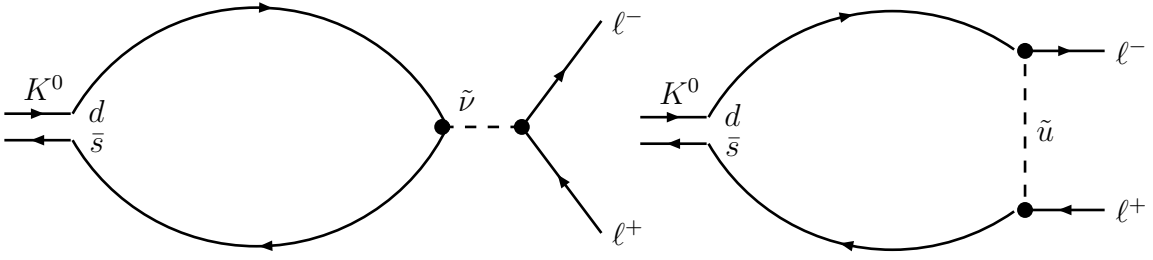

Two-body $K^0$ decays to muons and electrons proceed
via the diagrams shown in Fig.~\ref{fig:kaon0}.

At the quark level, the effective Lagrangian for such processes
has the form~\footnote{Note that, for energies of
the order of the kaon mass, both $s$- and $t$-channel
diagrams yield contact interactions.}
\cite{RviolK}:
\begin{eqnarray}
{\cal L}_{d\bar s\to \ell^-_j \ell^+_k}&=&
\frac{1}{m^2_{\tilde \nu_i}}\left[
\lambda^*_{ijk}\lambda'_{i12}
\left(\overline{s_R}d_L\right)\left(\overline{{\ell_j}_L}{\ell_k}_R\right)+
\lambda_{ikj}\lambda'^*_{i21}
\left(\overline{s_L}d_R\right)\left(\overline{{\ell_j}_R}{\ell_k}_L\right)
\right]
\nonumber\\
&-&{\lambda'^*_{ji1}\lambda'_{ki2}\over 2 m^2_{\tilde u_i}}
\left(\overline{s_R}\gamma_\mu d_R\right)
\left(\overline{{\ell_j}_L}\gamma^\mu{\ell_k}_L\right).
\label{quark}
\end{eqnarray}
The two different contributions
from $s$-channel and $t$-channel diagrams put
limits on different couplings~\cite{RviolK},
as we discuss below.

We have derived Feynman rules for the relevant effective
kaon, pion and lepton interactions. Based on these Feynman rules, we have 
recalculated the important kaon decay processes,
and update the limits on the products of  $R$-violating couplings
using the present experimental limits (\ref{Kprocesses}).  
The diagrams of Fig.~\ref{fig:kaon0} lead to the  the following effective
Lagrangian for $K^0 \ell^+  \ell^-$ interactions:
\begin{eqnarray}
{\cal L}_{K^0  \ell^-_j \ell^+_k}&=&
{F_{K^0}\over 2 m_{\tilde\nu_i}^2}\left[
\lambda^*_{ijk} \lambda'_{i12}\left(\overline{{\ell_j}_L}{\ell_k}_R\right)-
\lambda_{ikj} \lambda'^*_{i21}\left(\overline{{\ell_j}_R}{\ell_k}_L\right)
\right]K^0(p_K)
\nonumber \\
&-&{f_K\over 4 m_{\tilde u_i}^2} \lambda'^*_{ji1}\lambda'_{ki2}
~p_K^\mu \left(\overline{{\ell_j}_L}\gamma_\mu {\ell_k}_L\right)K^0(p_K),
\label{eff2}
\end{eqnarray}
where $F_{K^0}=m_{K^0}^2 f_{K}/(m_s+m_d)$, 
$m_s+m_d \simeq 0.15$ GeV is the sum of the
current masses of the $s$ and $d$ quarks, and
$f_K=0.1598$ GeV is the kaon decay constant.
The value of $F_{K^0}$ is related to the
pseudoscalar $<0|\bar s\gamma^5 d|K^0>=-F_{K^0}$ matrix element,
and is obtained from $f_K$ by 
using the Dirac equations for quarks. 
All QCD corrections are included in this
phenomenological approach. In the following, we assume that the
$R$-violating couplings are real and that
only one of
their products in (\ref{quark}) is non-zero.

We have implemented the Feynman rules in the CompHEP
package~\cite{comphep}, using
the effective Lagrangian~(\ref{eff2}),
and have obtained the following results:
\begin{equation}
\Gamma_{K^0\to \ell^-_j \ell^+_k}(\tilde \nu_i )=
\frac{
(\lambda_{ijk}\lambda'_{i12})^2 F_{K^0}^2}{64\pi~m_{\tilde\nu_i}^4
m_{K^0}}
\left(1-\frac{m_{\ell_j}^2+m_{\ell_k}^2}{m_{K^0}^2}\right)
\Delta(m_{K^0},m_{\ell_j},m_{\ell_k})
\label{nu}
\end{equation}

\begin{equation}
\Gamma_{K^0\to \ell^-_j \ell^+_k}(\tilde u_i )=
\frac{(\lambda'_{ji1}\lambda'_{ki2})^2 f_K^2}
{256\pi~m_{\tilde u_i}^4 m_{K^0}}
\left(m_{\ell_j}^2+m_{\ell_k}^2-\frac{(m_{\ell_j}^2-m_{\ell_k}^2)^2}{m_{K^0}^2}\right)
\Delta(m_{K^0},m_{\ell_j},m_{\ell_k})
\end{equation}
where
$\Delta(a,b,c)=\sqrt{\left[a^2-(b+c)^2\right]\left[a^2-(b-c)^2\right]}$ 
is the triangle function. As there exist two similar contributions in
the $s$ channel, coming from terms with different couplings
$\lambda_{ijk}\lambda'_{i12}$ and $\lambda_{ikj}\lambda'_{i21}$,
as seen in the first line in
(\ref{eff2}), we give only one of them in (\ref{nu}).

We obtain the following nominal numerical results for ${K^0\to
\ell^+\ell^-}$
decay via $\tilde\nu$ exchange, and the corresponding limits on
the $\lambda\lambda'$
products (for the numerical results for $\Gamma_{K^0\to
\ell^+\ell^-}$
we have used the nominal values $\lambda=\lambda'=1$  and 
$ m_{\tilde\nu(\tilde u)}=100\mbox{~GeV}$):
\begin{eqnarray}
\hspace*{-2cm}\ &&\Gamma_{K^0\to e^+\mu^-}=1.57\times 10^{-12}
\mbox{~GeV},\  \lambda_{i21}\lambda'_{i12}
\times \left(\frac{100~\mbox{~GeV}}{m_{\tilde \nu}}\right)^2 \le 6.2
\times 10^{-9} \nonumber\\
\hspace*{-2cm}\ && \hspace*{5.6cm} \
\lambda_{i12}\lambda'_{i21}
\times \left(\frac{100~\mbox{~GeV}}{m_{\tilde \nu}}\right)^2 \le 6.2
\times 10^{-9} \nonumber\\
\hspace*{-2cm}\ &&\Gamma_{K^0\to e^-\mu^+} =1.57\times
10^{-12}\mbox{~GeV},\  \lambda_{i12}\lambda'_{i12}
\times \left(\frac{100~\mbox{~GeV}}{m_{\tilde \nu}}\right)^2 \le 6.2
\times 10^{-9}  \nonumber\\
\hspace*{-2cm}\ && \hspace*{5.6cm} \
\lambda_{i21}\lambda'_{i21}
\times \left(\frac{100~\mbox{~GeV}}{m_{\tilde \nu}}\right)^2 \le 6.2
\times 10^{-9} \nonumber\\
\hspace*{-2cm}\ &&\Gamma_{K^0\to e^+e^-}=1.72\times 10^{-12}\mbox{~GeV},\  
\lambda_{i11}\lambda'_{i12}
\times \left(\frac{100~\mbox{~GeV}}{m_{\tilde \nu}}\right)^2 \le 1.0
\times 10^{-8}   \nonumber\\
\hspace*{-2cm}\ && \hspace*{5.6cm} \
\lambda_{i11}\lambda'_{i21}
\times \left(\frac{100~\mbox{~GeV}}{m_{\tilde \nu}}\right)^2 \le 1.0
\times 10^{-8}   \nonumber\\
\hspace*{-2cm}\ &&\Gamma_{K^0\to \mu^+\mu^-}=1.42\times
10^{-12}\mbox{~GeV},\  \lambda_{i22}\lambda'_{i12}
\times \left(\frac{100~\mbox{~GeV}}{m_{\tilde \nu}}\right)^2 
\le 2.6\times 10^{-7}
\nonumber\\
\hspace*{-2cm}\ && \hspace*{5.6cm} \
\lambda_{i22}\lambda'_{i21}
\times \left(\frac{100~\mbox{~GeV}}{m_{\tilde \nu}}\right)^2
\le 2.6\times 10^{-7}
\label{res-k1}
\end{eqnarray}
For ${K^0\to \ell^+\ell^-}$ decay 
via up-squark exchange we have the following limits:
\begin{eqnarray}
\hspace*{-2cm}\ &&\Gamma_{K^0\to e^+\mu^-}=1.61\times
10^{-15}\mbox{~GeV},\  
\lambda'_{2i1}\lambda'_{1i2}
\times \left(\frac{100~\mbox{~GeV}}{m_{\tilde u}}\right)^2 \le 1.9
\times 10^{-7} \nonumber\\
\hspace*{-2cm}\ &&\Gamma_{K^0\to e^-\mu^+}=1.61\times
10^{-15}\mbox{~GeV},\  
\lambda'_{1i1}\lambda'_{2i2}
\times \left(\frac{100~\mbox{~GeV}}{m_{\tilde u}}\right)^2 \le 1.9
\times 10^{-7} \nonumber\\
\hspace*{-2cm}\ &&\Gamma_{K^0\to e^+e^-}=8.25\times 10^{-20}\mbox{~GeV},\  
\lambda'_{1i1}\lambda'_{1i2}
\times \left(\frac{100~\mbox{~GeV}}{m_{\tilde u}}\right)^2 \le 4.7
\times 10^{-5} \nonumber\\
\hspace*{-2cm}\ &&\Gamma_{K^0\to \mu^+\mu^-}=3.19\times
10^{-15}\mbox{~GeV},\  
\lambda'_{2i1}\lambda'_{2i2}
\times \left(\frac{100~\mbox{~GeV}}{m_{\tilde u}}\right)^2 \le 5.4
\times 10^{-6}
\label{res-k2}
\end{eqnarray}
We have used in our calculations the decay width 
$\Gamma_{exp}(K^0_L)=1.273\times 10^{-17}$ GeV
and the experimental limits on $K^0\to \ell^+\ell^-$ decay 
widths shown in (\ref{Kprocesses}).

\ \\

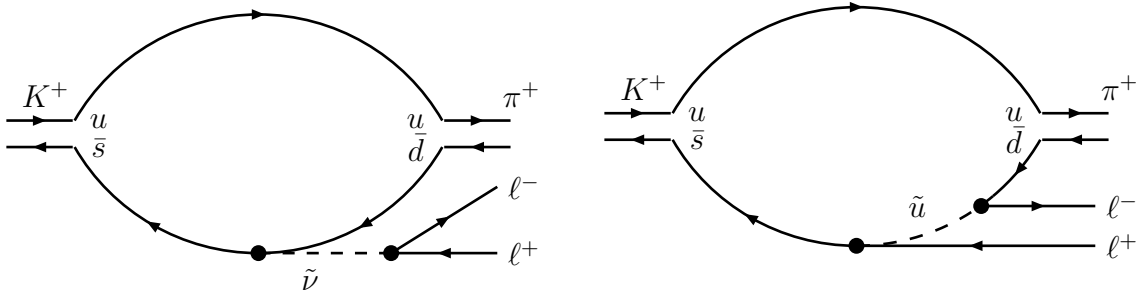
\begin{figure}[h]
\hspace*{-1cm}
\begin{picture}(220,110)(0,0)
\SetWidth{1.}      
\ArrowLine(40,35)(15,35)    \ArrowLine(15,45)(40,45) \Text(30,55)[]{$K^+$}
\ArrowLine(205,35)(180,35) \ArrowLine(180,45)(205,45)\Text(210,55)[]{$\pi^+$}
\ArrowArcn(110,5 )(80,150,30) \ArrowArcn(110,75)(80,-90,-150)
\ArrowArcn(110,75)(80,-30,-90)
\DashLine(110,-5)(160,-5){5}
\Vertex(110,-5){3}
\Text(130,-15)[]{$\tilde\nu$}
\ArrowLine(200,-5)(160,-5)\Vertex(160,-5){3}\Text(205,-5)[l]{$\ell^+$}
\ArrowLine(160,-5)(200,20)\Text(205,20)[l]{$\ell^-$}
\Text(50,45)[]{$u$}
\Text(50,35)[]{$\bar s$}
\Text(170,35)[]{$\bar d$}
\Text(170,45)[]{$ u$}
\end{picture}

\vspace*{-4cm}
\hspace*{7.cm}
\begin{picture}(220,110)(0,0)
\SetWidth{1.}      
\ArrowLine(40,35)(15,35)    \ArrowLine(15,45)(40,45) \Text(30,55)[]{$K^+$}
\ArrowLine(205,35)(180,35) \ArrowLine(180,45)(205,45)\Text(210,55)[]{$\pi^+$}
\ArrowArcn(110,5 )(80,150,30) \ArrowArcn(110,75)(80,-90,-150) 
\ArrowArcn(110,75)(80,-30,-50) \DashCArc(110,75)(80,-90,-50){5}
\Text(130,10)[l]{$\tilde u$}
\Text(50,35)[]{$\bar s$}
\Text(50,45)[]{$u$}
\Text(170,45)[]{$ u$}
\Text(170,35)[]{$\bar d$}
\ArrowLine(200,-5)(110,-5)\Vertex(110,-5){3}\Text(205,-5)[l]{$\ell^+$}
\ArrowLine(157,10)(200,10)\Text(205,10)[l]{$\ell^-$}
\Vertex(157,10){3}
\end{picture}

\vspace*{0.5cm}

\caption{\label{fig:kaon1}
\it Diagrams  involving $R$-violating couplings that yield the three-body
leptonic decays $K^+ \to\pi^+ \ell^{-}\ell^{+}$.}
\end{figure}

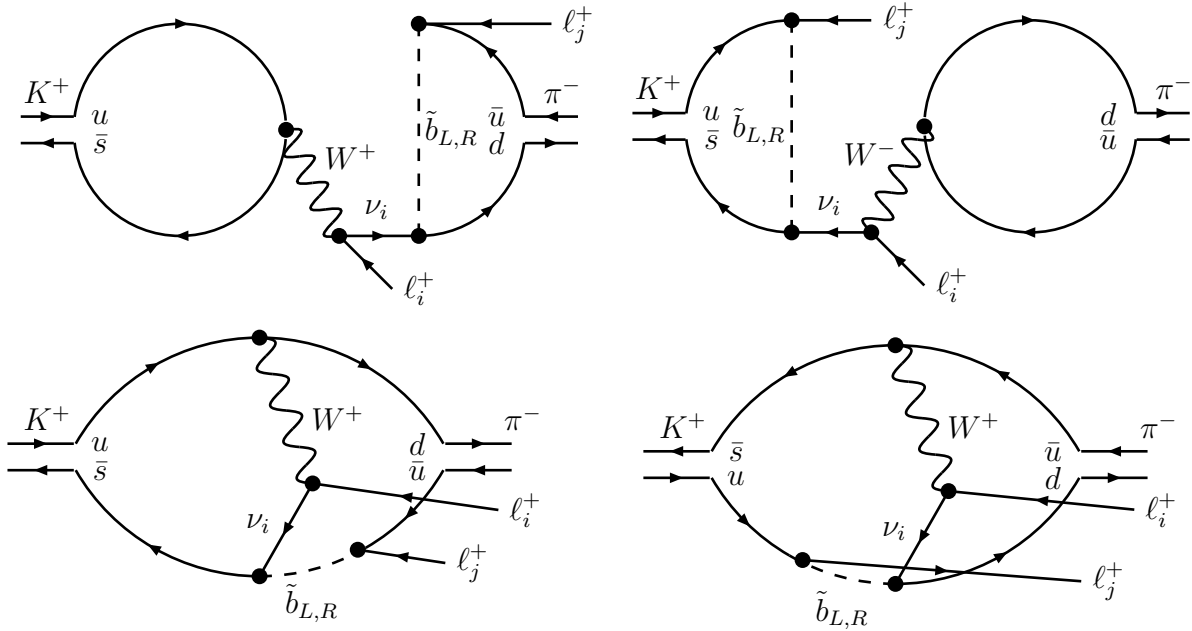
\begin{figure}[h]
\vspace*{-1cm}
\hspace*{-1.5cm}
\begin{picture}(220,120)(0,0)
\SetWidth{1.}      
\ArrowLine(40,35)(20,35) \ArrowLine(20,45)(40,45) \Text(30,55)[]{$K^+$}
\ArrowArcn(80,40 )(40,173,5) \ArrowArcn(80,40)(40,-5,-173)
\Text(50,45)[]{$u$}
\Text(50,35)[]{$\bar s$}
\Photon(120,40)(140,0){5}{4}\Text(135,30)[l]{$W^+$}
\Vertex(120,40){3}
\ArrowLine(160,-20)(140,0)\Text(165,-20)[l]{$\ell^+_i$}
\Vertex(140,0){3}
\ArrowLine(140,0)(170,0)\Text(150,10)[l]{$\nu_i$}
\Vertex(170,0){3}
\ArrowArc(170,40 )(40,-90,-7)
\ArrowArc(170,40 )(40,7,90)
\DashLine(170,0)(170,80){5}\Text(183,40)[]{$\tilde b_{L,R}$}
\Vertex(170,80){3}
\ArrowLine(220,80)(170,80)\Text(225,80)[l]{$\ell^+_j$}
\ArrowLine(230,45)(210,45) \ArrowLine(210,35)(230,35)
\Text(200,45)[]{$\bar u$}
\Text(200,35)[]{$d$}
\Text(225,55)[]{$\pi^-$}
\end{picture}

\vspace*{-4.3cm}  
\hspace*{8cm}
\hspace*{-1.5cm}  
\begin{picture}(220,120)(0,0)
\SetWidth{1.}      
\ArrowLine(40,35)(20,35) \ArrowLine(20,45)(40,45) \Text(30,55)[]{$K^+$}
\ArrowArcn(80,40 )(40,173,90) \ArrowArcn(80,40)(40,-90,-173)
\Text(50,45)[]{$u$}
\Text(50,35)[]{$\bar s$}
\DashLine(80,0)(80,80){5}\Text(68,40)[]{$\tilde b_{L,R}$} 
\Vertex(80,0){3}\Vertex(80,80){3}
\ArrowLine(110,0)(80,0)\Text(90,10)[l]{$\nu_i$}
\ArrowLine(110,80)(80,80)\Text(115,80)[l]{$\ell^+_j$}
\ArrowLine(130,-20)(110,0)
\Text(135,-20)[l]{$\ell^+_i$}
\Photon(110,0)(130,40){5}{4}\Text(100,30)[l]{$W^-$}
\Vertex(110,0){3}\Vertex(130,40){3}
\ArrowArcn(170,40 )(40,175,7) \ArrowArcn(170,40)(40,-7,-175)
\ArrowLine(210,45)(230,45) \ArrowLine(230,35)(210,35)
\Text(200,45)[]{$d$}
\Text(200,35)[]{$\bar u$}
\Text(225,55)[]{$\pi^-$}
\end{picture}

\vspace*{0.5cm}
\hspace*{-1.5cm}
\begin{picture}(220,110)(0,0)
\SetWidth{1.}      
\ArrowLine(40,35)(15,35)\ArrowLine(15,45)(40,45)\Text(30,55)[]{$K^+$}
\ArrowLine(205,35)(180,35)\ArrowLine(180,45)(205,45)\Text(210,55)[]{$\pi^-$}
\ArrowArcn(110,5 )(80,150,90) \ArrowArcn(110,5 )(80,90,30)
\ArrowArcn(110,75)(80,-90,-150)
\ArrowArcn(110,75)(80,-30,-60) \DashCArc(110,75)(80,-90,-60){5}
\Text(120,-15)[l]{$\tilde b_{L,R}$}
\Text(50,45)[]{$u$}
\Text(50,35)[]{$\bar s$}
\Text(170,45)[]{$d$}
\Text(170,35)[]{$\bar u$}
\Vertex(147,5){3}
\Vertex(110,85){3}
\Photon(110,85)(130,30){5}{4}\Text(140,55)[]{$W^+$}
\Vertex(130,30){3}
\ArrowLine(130,30)(110,-5) \Text(110,15)[]{$\nu_i$}  
\Vertex(110,-5){3}
\ArrowLine(200,20)(130,30)  \Text(205,20)[l]{$\ell^+_i$}
\ArrowLine(180,0)(147,5) \Text(185,0)[l]{$\ell^+_j$}
\end{picture}

\vspace*{-3.8cm}
\hspace*{7.cm}
\begin{picture}(220,110)(0,0)
\SetWidth{1.}      
\ArrowLine(15,35)(40,35)\ArrowLine(40,45)(15,45)   \Text(30,55)[]{$K^+$}
\ArrowLine(180,35)(205,35)
\ArrowLine(205,45)(180,45)\Text(210,55)[]{$\pi^-$}
\ArrowArc(110,5 )(80,90,150)\ArrowArc(110,5 )(80,30,90)
\ArrowArc(110,75)(80,-150,-120)
\ArrowArc(110,75)(80,-90,-30) \DashCArc(110,75)(80,-120,-90){5}
\Text(80,-15)[l]{$\tilde b_{L,R}$}
\Text(50,45)[]{$\bar s$} 
\Text(50,35)[]{$u$}
\Text(170,45)[]{$\bar u$}
\Text(170,35)[]{$d$}
\Photon(110,85)(130,30){5}{4}\Text(140,55)[]{$W^+$}
\ArrowLine(130,30)(110,-5) \Text(110,15)[]{$\nu_i$}
\Vertex(130,30){3}\Vertex(110,-5){3}
\ArrowLine(200,23)(130,30)  \Text(205,23)[l]{$\ell^+_i$}
\Vertex(110,85){3}
\ArrowLine(75,4)(180,-4) \Text(185,-4)[l]{$\ell^+_j$}
\Vertex(75,4){3}
\end{picture}
\vspace*{0.5cm}
\caption{\label{fig:kaon2}
\em Diagrams involving $R$-violating couplings that yield three-body
like-sign leptons decays $K^+\to\pi^- \ell^+_i\ell^+_j$.}
\end{figure}

We now discuss the diagrams for 3-body kaon
decays to pions and two charged leptons, of which
there are two qualitatively different kinds:

\begin{itemize}

\item
The kaon may decay into a pion of the same charge, in which case the
leptons in the final state must have opposite signs:  
$K^\pm\to\pi^\pm \ell^\mp\ell'^\pm$
and $K^0\to\pi^0 \ell^\mp\ell'^\pm$. The corresponding
diagram for the first process is shown in Fig.~\ref{fig:kaon1}.

\item
The kaon may decay into a pion with the opposite charge, in which
case the leptons in the final state must have the same signs: 
$K^\pm\to\pi^\mp \ell^\pm\ell'^\pm$.
Representative diagrams for this process are shown in
Fig.~\ref{fig:kaon2}.
This process involves two heavy virtual  particles, the $W$ boson and a
down squark. One should note that  decay width of this process is 
directly proportional
to the mixing between the left- and right-handed squark states,
denoted by $\tilde b_L$  and $\tilde b_R$, respectively.
If there is no mixing, this same-sign-lepton
process mentioned vanishes. One can expect sizeable
mixing  only for squarks of the third generation (and especially in
the high-$\tan\beta$
region), which is why we have used $\tilde b_{L,R}$ in the diagram.  

\end{itemize}

The possibilities for 
$K^\pm\to\pi^\pm  \ell^\mp\ell'^\pm$ or 
$K^0\to\pi^0  \ell^\mp\ell'^\pm$ decay may
be further subdivided into two groups.
\begin{itemize}
\item
Diagrams involving only squarks, via which experimental upper limits
bound products of $LQ\bar{D}$ operators.

\item
Diagrams involving also sleptons, which yield bounds on products of 
$LQ\bar{D}$ and $LL\bar{E}$ operators.
\end{itemize}

The diagrams of Fig.~\ref{fig:kaon1} lead to the  the following effective
Lagrangian for $K^+ \pi^- \ell^+  \ell^-$ interactions:

\begin{eqnarray}
{\cal L}_{K^+ \pi^- \ell^-_k \ell^+_j  } &=& 
{M_s\over m_{\tilde\nu_i}^2}\left[
\lambda^*_{ijk} \lambda'_{i12}\left(\overline{{\ell_j}_L}{\ell_k}_R\right)+
\lambda_{ikj} \lambda'^*_{i21}\left(\overline{{\ell_j}_R}{\ell_k}_L\right)
\right]K^+(p_K)\pi^-(-p_\pi)
\nonumber \\
&+&{f_+\over 4 m_{\tilde u_i}^2}
\lambda'^*_{ji1}\lambda'_{ki2}
\left(p_K+p_\pi\right)^\mu 
\left(\overline{{\ell_j}_L}\gamma_\mu {\ell_k}_L\right)K^+(p_K)\pi^-(-p_\pi),
\label{3body}
\end{eqnarray}
where $M_s \simeq 0.49$ GeV is the constituent mass of the $s$ quark.
In the general case, the vector matrix element of the $K \rightarrow \pi$ 
transition is parametrized by two momentum-dependent
form factors $f_+(t)$ and $f_-(t)$: 
\begin{eqnarray}
&&<0|\bar s\gamma^\mu d|K^+(p_K)\pi^-(-p_\pi)>=
\nonumber\\
&&<\pi^+(p_\pi)|\bar s\gamma^\mu d|K^+(p_K)>=
(p_K+p_\pi)^\mu f_+(t)+(p_K-p_\pi)^\mu f_-(t),
\nonumber
\end{eqnarray}
where $t=(p_K-p_\pi)^2$ is the squared momentum transfered to the lepton pair.
The experimental data on semileptonic kaon decay
are adequately described by a linear approximation
for $f_+$:
$$
f_+(t)=f_+(0)\left[1+\lambda_+(t/m^2_\pi)\right],
$$
where $\lambda_+\simeq 0.03$ and $f_- \simeq -0.31\pm0.15$. 
In the case of exact flavour $SU(3)$ symmetry, the following relations
hold: $f_+(0)=1$ and $f_-(0)=0$. Due to the Ademollo-Gatto theorem, the 
first form factor is renormalized only at second order in the 
$SU(3)$-violating interactions, and is therefore expected to be close to
unity. The Ademollo-Gatto theorem is not applicable to $f_-$, whose
value is close to zero, as follows from experiments and the 
Callan-Treiman relation:
$$
f_-(m^2_K)=\frac{f_K}{f_\pi}-f_+(m^2_K)\simeq -0.15~.
$$
For our estimations, we set $f_+(0)=1$ and neglect $f_-$.
To estimate the scalar form factor, we use the relativistic quark model,
which gives~\cite{ScFormFact}:
$$
<0|\bar s d|K^+(p_K)\pi^-(-p_\pi)>=<\pi^+(p_\pi)|\bar s d|K^+(p_K)>
\simeq-2 M_s,
$$
where we keep only the leading term and drop the momentum dependence,
and $M_s\simeq 0.49$ GeV as before.
In the approximation of unbroken $SU(3)$ symmetry,
the corresponding form factors for the neutral kaons are smaller
by a factor of $\sqrt{2}$.

We can now estimate the decay rates of the charged kaons:
\begin{eqnarray}
\Gamma_{K^+\to \pi^+\ell^-_j \ell^+_k}(\tilde \nu_i)=
\frac{
(\lambda_{ijk}\lambda'_{i12})^2 M_s^2}{256\pi^3~m_{\tilde\nu_i}^4
m^3_{K^+}}
\int\limits_{(m_{\ell_j}+m_{\ell_k})^2}^{(m_{K^+}-m_{\pi^+})^2}
\!\!\!\!\!\!\!\!\!&&
\Delta(t,m_{\ell_j},m_{\ell_k})\Delta(t,m_{K^+},m_{\pi^+})
\nonumber
\\
&&\times\left(t^2-m_{\ell_j}^2-m_{\ell_k}^2\right)\frac{\mbox{d}t^2}{t^2},
\end{eqnarray}

\begin{eqnarray}
\Gamma_{K^+\to \pi^+\ell^-_j \ell^+_k}(\tilde u_i )&=&
\frac{(\lambda'_{ji1}\lambda'_{ki2})^2}
{2048\pi^3~m_{\tilde u_i}^4 m^3_{K^+}}
\int\limits_{(m_{\ell_j}+m_{\ell_k})^2}^{(m_{K^+}-m_{\pi^+})^2}
\!\!\!\!\!\!\!\!\!
\Delta(t,m_{\ell_j},m_{\ell_k})\Delta(t,m_{K^+},m_{\pi^+})
\nonumber
\\
&&\times\left[
\left(m^2_{K^+}+m^2_{\pi^+}+m^2_{\ell_j}+m^2_{\ell_k}-t^2\right)~F_1(t) 
- F_2(t)\right.
\nonumber
\\
&&\left. ~~~+ 
\frac{m^2_{\ell_j}+m^2_{\ell_k}}{2}\left(t^2-m_{\ell_j}^2-m_{\ell_k}^2\right)
-2m^2_{\pi^+} m^2_{K^+}
\right]
\nonumber
\\
&&\times\left[1+\lambda_+(t^2/m^2_\pi)\right]^2
\frac{\mbox{d}t^2}{t^2},
\end{eqnarray}
\noindent
where $F_1(t)=q_{max}(t)+q_{min}(t)$, 
$F_2(t)=2(q^2_{max}(t)+q_{max}(t)q_{min}(t)+q^2_{min}(t))/3$, and
\begin{eqnarray}
q_{max}(t)=\frac{1}{4t^2}
\left[\left(m^2_{K^+}-m^2_{\pi^+}+m^2_{\ell_j}-m^2_{\ell_k}\right)^2-
\left(\Delta(t,m_{\ell_j},m_{\ell_k})-
\Delta(t,m_{K^+},m_{\pi^+})\right)^2\right],
\nonumber\\
q_{min}(t)=\frac{1}{4t^2}  
\left[\left(m^2_{K^+}-m^2_{\pi^+}+m^2_{\ell_j}-m^2_{\ell_k}\right)^2-           
\left(\Delta(t,m_{\ell_j},m_{\ell_k})+
\Delta(t,m_{K^+},m_{\pi^+})\right)^2\right].
\nonumber
\end{eqnarray}
For the decay rates of the neutral kaons, the $SU(3)$ factor of 1/2 and 
the corresponding masses for the neutral mesons $m_{K^0}$ and $m_{\pi^0}$
must be taken into account.

We obtain the following nominal numerical results for 
${K^+\to \pi^+ \ell^+\ell^-}$ and ${K^0\to \pi^0 \ell^+\ell^-}$
decays via $\tilde\nu$ exchange, and the corresponding limits on the
$\lambda\lambda'$
products (for the numerical results
we have used the nominal values $\lambda=\lambda'=1$  and 
$ m_{\tilde\nu(\tilde u)}=100\mbox{~GeV}$):

\begin{eqnarray}
\hspace*{-2cm}\ &&\Gamma_{K^+\to \pi^+ e^+\mu^-}=1.38\times 10^{-15} 
\mbox{~GeV},\  
\lambda_{i21}\lambda'_{i12}
\times \left(\frac{100~\mbox{~GeV}}{m_{\tilde \nu}}\right)^2 \le 4.5
\times 10^{-6} \nonumber\\
\hspace*{-2cm}\ && \hspace*{6cm} \
\lambda_{i12}\lambda'_{i21}
\times \left(\frac{100~\mbox{~GeV}}{m_{\tilde \nu}}\right)^2 \le 4.5
\times 10^{-6} \nonumber\\
\hspace*{-2cm}\ &&\Gamma_{K^0\to \pi^0 e^+\mu^-}=7.71\times 10^{-16}
\mbox{~GeV},\
\lambda_{i21}\lambda'_{i12}
\times \left(\frac{100~\mbox{~GeV}}{m_{\tilde \nu}}\right)^2 \le 7.2
\times 10^{-6} \nonumber\\
\hspace*{-2cm}\ && \hspace*{6cm} \
\lambda_{i12}\lambda'_{i21}
\times \left(\frac{100~\mbox{~GeV}}{m_{\tilde \nu}}\right)^2 \le 7.2
\times 10^{-6} \nonumber\\
\hspace*{-2cm}\ &&\Gamma_{K^+\to \pi^+ e^-\mu^+}=1.38\times 10^{-15}
\mbox{~GeV},\
\lambda_{i12}\lambda'_{i12}
\times \left(\frac{100~\mbox{~GeV}}{m_{\tilde \nu}}\right)^2 \le 1.0
\times 10^{-6} \nonumber\\
\hspace*{-2cm}\ && \hspace*{6cm} \
\lambda_{i21}\lambda'_{i21}
\times \left(\frac{100~\mbox{~GeV}}{m_{\tilde \nu}}\right)^2 \le 1.0
\times 10^{-6} \nonumber\\
\hspace*{-2cm}\ &&\Gamma_{K^0\to \pi^0 e^-\mu^+}=7.71\times 10^{-16}
\mbox{~GeV},\
\lambda_{i12}\lambda'_{i12}
\times \left(\frac{100~\mbox{~GeV}}{m_{\tilde \nu}}\right)^2 \le 7.2
\times 10^{-6} \nonumber\\
\hspace*{-2cm}\ && \hspace*{6cm} \
\lambda_{i21}\lambda'_{i21}
\times \left(\frac{100~\mbox{~GeV}}{m_{\tilde \nu}}\right)^2 \le 7.2
\times 10^{-6} \nonumber\\
\hspace*{-2cm}\ &&\Gamma_{K^+\to\pi^+e^+e^-} =2.14\times
10^{-15}\mbox{~GeV},\  
\lambda_{i11}\lambda'_{i12}
\times \left(\frac{100~\mbox{~GeV}}{m_{\tilde \nu}}\right)^2 \le 8.8
\times 10^{-5}   \nonumber\\
\hspace*{-2cm}\ && \hspace*{6cm} \
\lambda_{i11}\lambda'_{i21}
\times \left(\frac{100~\mbox{~GeV}}{m_{\tilde \nu}}\right)^2 \le 8.8
\times 10^{-5}   \nonumber\\
\hspace*{-2cm}\ &&\Gamma_{K^0\to\pi^0e^+e^-} =1.17\times
10^{-15}\mbox{~GeV},\
\lambda_{i11}\lambda'_{i12}
\times \left(\frac{100~\mbox{~GeV}}{m_{\tilde \nu}}\right)^2 \le 6.8
\times 10^{-6}   \nonumber\\
\hspace*{-2cm}\ && \hspace*{6cm} \
\lambda_{i11}\lambda'_{i21}
\times \left(\frac{100~\mbox{~GeV}}{m_{\tilde \nu}}\right)^2 \le 6.8
\times 10^{-6}   \nonumber\\
\hspace*{-2cm}\ &&\Gamma_{K^+\to \pi^+\mu^+\mu^-}=7.58\times 10^{-16}
\mbox{~GeV},\  
\lambda_{i22}\lambda'_{i12}
\times \left(\frac{100~\mbox{~GeV}}{m_{\tilde \nu}}\right)^2 \le 8.2
\times 10^{-5}   \nonumber\\
\hspace*{-2cm}\ && \hspace*{6cm} \
\lambda_{i22}\lambda'_{i21}
\times \left(\frac{100~\mbox{~GeV}}{m_{\tilde \nu}}\right)^2 \le 8.2
\times 10^{-5}   \nonumber\\
\hspace*{-2cm}\ &&\Gamma_{K^0\to \pi^0\mu^+\mu^-}=4.38\times 10^{-16}
\mbox{~GeV},\
\lambda_{i22}\lambda'_{i12}
\times \left(\frac{100~\mbox{~GeV}}{m_{\tilde \nu}}\right)^2 \le 1.2
\times 10^{-5}   \nonumber\\
\hspace*{-2cm}\ && \hspace*{6cm} \
\lambda_{i22}\lambda'_{i21}
\times \left(\frac{100~\mbox{~GeV}}{m_{\tilde \nu}}\right)^2 \le 1.2
\times 10^{-5}
\end{eqnarray}

For the decays 
via up-squark exchange we have the following limits:
\begin{eqnarray}
\hspace*{-2cm}\ &&\Gamma_{K^+\to \pi^+e^+\mu^-}=1.61\times 10^{-16}
\mbox{~GeV},\  
\lambda'_{2i1}\lambda'_{1i2}
\times \left(\frac{100~\mbox{~GeV}}{m_{\tilde u}}\right)^2 \le 1.3
\times 10^{-5} \nonumber\\
\hspace*{-2cm}\ &&\Gamma_{K^0\to \pi^0e^+\mu^-}=9.06\times 10^{-17}
\mbox{~GeV},\   
\lambda'_{2i1}\lambda'_{1i2}
\times \left(\frac{100~\mbox{~GeV}}{m_{\tilde u}}\right)^2 \le 2.1
\times 10^{-5} \nonumber\\
\hspace*{-2cm}\ &&\Gamma_{K^+\to \pi^+ e^-\mu^+}=1.61\times 10^{-16}
\mbox{~GeV},\  
\lambda'_{1i1}\lambda'_{2i2}
\times \left(\frac{100~\mbox{~GeV}}{m_{\tilde u}}\right)^2 \le 3.0
\times 10^{-6} \nonumber\\
\hspace*{-2cm}\ &&\Gamma_{K^0\to \pi^0 e^-\mu^+}=9.06\times 10^{-17}
\mbox{~GeV},\
\lambda'_{1i1}\lambda'_{2i2}
\times \left(\frac{100~\mbox{~GeV}}{m_{\tilde u}}\right)^2 \le 2.1
\times 10^{-5} \nonumber\\
\hspace*{-2cm}\ &&\Gamma_{K^+\to \pi^+e^+e^-}=2.38\times
10^{-16}\mbox{~GeV},\  
\lambda'_{1i1}\lambda'_{1i2}
\times \left(\frac{100~\mbox{~GeV}}{m_{\tilde u}}\right)^2 \le 2.7
\times 10^{-4} \nonumber\\
\hspace*{-2cm}\ &&\Gamma_{K^0\to \pi^0e^+e^-}=1.31\times
10^{-16}\mbox{~GeV},\
\lambda'_{1i1}\lambda'_{1i2}
\times \left(\frac{100~\mbox{~GeV}}{m_{\tilde u}}\right)^2 \le 2.0
\times 10^{-5} \nonumber\\
\hspace*{-2cm}\ &&\Gamma_{K^+\to\pi^+ \mu^+\mu^-}=9.33\times 10^{-17}
\mbox{~GeV},\  
\lambda'_{2i1}\lambda'_{2i2}
\times \left(\frac{100~\mbox{~GeV}}{m_{\tilde u}}\right)^2 \le 2.3
\times 10^{-4} \nonumber\\
\hspace*{-2cm}\ &&\Gamma_{K^0\to\pi^0 \mu^+\mu^-}=5.48\times 10^{-17}
\mbox{~GeV},\
\lambda'_{2i1}\lambda'_{2i2}
\times \left(\frac{100~\mbox{~GeV}}{m_{\tilde u}}\right)^2 \le 3.4
\times 10^{-5}
\end{eqnarray}

We have used in our calculations the decay width 
$\Gamma_{exp}(K^+)=5.314\times 10^{-17}$ GeV
and the present limits on $K^+\to \pi^+\ell^+\ell^-$ 
and $K_L^0\to \pi^0\ell^+\ell^-$ decay widths~(\ref{Kprocesses}).
Note that there are different constraints from 
the $K^+\to \pi^+ e^+\mu^-$ and $K^+\to \pi^+e^-\mu^+$ decays, 
because of the rather different experimental limits $BR(K^+\to \pi^+
e^+\mu^-)\le 6.9\times 10^{-9}$ and
$BR(K^+\to \pi^+ e^-\mu^+)\le 2.8\times 10^{-11}$.
The limit obtained from $K^0\to\ell^+\ell^-$ is typically 1-2 orders 
of magnitude better than that derived from $K^+\to \pi^+\ell^+\ell^-$ decay.

As already mentioned,
diagrams with non-zero sbottom-quark mixing
may lead to like-sign leptons, as seen in Fig.~\ref{fig:kaon2}.
They arise from the effective lepton-number violating contact 
interactions
\begin{equation}
{\cal L}=
-\frac{\lambda'_{ikp}\lambda'_{jqk}V_{LR}}{m^2_{d_k}}
\left(\overline{{d_p}_R}{\nu_i}_L\right)
\left(\overline{({\ell_j}_L)^C}{u_q}_L\right)+ \mbox{h.c.},
\end{equation}
where $V_{LR}$ denotes left-right squark mixing matrix element.

There are two different topologies, shown in the first and second row
of Fig.~\ref{fig:kaon2}, respectively. Diagrams in the first row lead to
an
effective tree-level process, since the $W$-boson 
and squark masses
are much bigger then the typical energy scale, which is of the order of
the kaon mass.
Diagrams in the second row cannot be reduced to tree-level diagrams,
since one
has a neutrino propagator in the loop. However, this last set of
diagrams give a contribution that is typically
about 2 orders of magnitude lower than the  diagrams of the 
first row~\cite{double-beta}. Therefore, for the sake of simplicity, we
neglect the second-row diagrams,
and have derived an effective Lagrangian only for the first two  diagrams 
of Fig.~\ref{fig:kaon2}.
Two kinds of effective interaction appear:
Standard-Model-like $K \ell\nu$ and $\pi \ell\nu$ interactions
and new effective interactions related to $R$-violating operators of
the forms $K \ell^C\nu$ and $\pi \ell^C\nu$.

The effective
Lagrangian for those interactions take the following forms:

\begin{equation}
{\cal L}_{K^+ \ell^-_i \bar\nu_i } = 
V_{us} \sqrt{2} G_F f_K p_K^\mu 
\left(\overline{{\nu_i}_L} \gamma_\mu {\ell_i}_L \right)K^+(p_K)
\end{equation}
\begin{equation}
{\cal L}_{\pi^+ \ell^-_i \bar\nu_i } = 
V_{ud} \sqrt{2} G_F f_\pi p_\pi^\mu 
\left(\overline{{\nu_i}_L} \gamma_\mu {\ell_i}_L \right)\pi^+(p_\pi)
\end{equation}
where $f_{K}$ and $f_{\pi}=0.1307$ GeV 
are the kaon and pion decay constants, respectively,
$G_F=1.16639 10^{-5}$ GeV$^{-2}$ is the Fermi constant 
and  $V_{us},V_{ud} $ are CKM matrix elements, and:

\begin{equation}
{\cal L}_{K^+ (\ell^-_j)^C \nu_i } (\tilde d_k) = 
\frac{\lambda'_{ik2}\lambda'_{j1k} V_{LR}}{4 m_{\tilde d_k}^2}
 F_{K^+} \left(\overline{({\ell_j}_L)^C} {\nu_i}_L\right)K^+(p_K)
\end{equation}

\begin{equation}
{\cal L}_{\pi^+ (\ell^-_j)^C \nu_i } (\tilde d_k) = 
\frac{\lambda'_{ik1}\lambda'_{j1k}V_{LR}}{4 m_{\tilde d_k}^2}
F_{\pi^+}\left(\overline{({\ell_j}_L)^C} {\nu_i}_L\right)\pi^+(p_\pi)
\end{equation}
where $F_{K^+}=m^2_{K^+}f_K/(m_s+m_u)$,
$F_{\pi^+}=m^2_{\pi^+}f_\pi/(m_d+m_u)$,
$m_s+m_u\simeq 0.15$ GeV, and $m_d+m_u\simeq 0.01$ GeV.

\begin{figure}[h]
{
\unitlength=1.0 pt
\SetScale{1.}
\SetWidth{1.0}      
{} \qquad\allowbreak
\begin{picture}(95,79)(0,0)
\Text(15.0,60.0)[r]{$K^+$}
\DashArrowLine(16.0,60.0)(58.0,60.0){1.0}
\Text(80.0,70.0)[l]{$\ell^+_i$}
\ArrowLine(79.0,70.0)(58.0,60.0)
\Text(54.0,50.0)[r]{$\nu_i$}
\ArrowLine(58.0,60.0)(58.0,40.0)
\Text(80.0,50.0)[l]{$\pi^-$}
\DashArrowLine(58.0,40.0)(79.0,50.0){1.0}
\Text(80.0,30.0)[l]{$\ell^+_j$}
\ArrowLine(79.0,30.0)(58.0,40.0)
\end{picture} \
{} \qquad\allowbreak \ \ \ \ \
\begin{picture}(95,79)(0,0)
\Text(15.0,60.0)[r]{$K^+$}
\DashArrowLine(16.0,60.0)(58.0,60.0){1.0}
\Text(80.0,70.0)[l]{$\ell^+_j$}
\ArrowLine(79.0,70.0)(58.0,60.0)   
\Text(54.0,50.0)[r]{$\nu_i$}
\ArrowLine(58.0,40.0)(58.0,60.0)
\Text(80.0,50.0)[l]{$\pi^-$}
\DashArrowLine(58.0,40.0)(79.0,50.0){1.0}
\Text(80.0,30.0)[l]{$\ell^+_i$}
\ArrowLine(79.0,30.0)(58.0,40.0)
\end{picture} \
}
\caption{\label{fig:kaon3}
\em Diagrams for the like-sign lepton decay
$K^+\to\pi^- \ell^+_i\ell^+_j$,
in terms of the effective Standard-Model-like interactions $K \ell\nu$ and
$\pi \ell\nu$ and effective $K \ell^C\nu$ and $\pi\ell^C\nu$ interactions
related to $R$-violating operators.}
\end{figure}
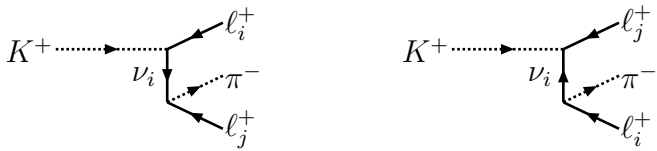

The Feynman diagrams for $K^+\to\pi^- \ell^+_i\ell^+_j$ decay 
are given in terms of these effective interactions, as shown in
Fig.~\ref{fig:kaon3}. They yield the matrix elements:
\begin{equation}
{\cal M} = {\lambda'_{ik1}\lambda'_{j1k}V_{LR}F_{\pi^+}\over 
2 m^2_{\tilde d_k}}
V_{us}\frac{G_F}{\sqrt{2}}f_K m_{\ell_i}
\left(\overline{({\ell_j}_L)^C}~{\hat p_{\nu_i}\over p^2_{\nu_i}}
{\ell_i}_R\right)K^+(p_K)\pi^-(p_\pi)
\end{equation}
and 
\begin{equation}
{\cal M} = {\lambda'_{ik2}\lambda'_{j1k}V_{LR}F_{K^+}\over 
2 m^2_{\tilde d_k}}
V_{ud}\frac{G_F}{\sqrt{2}}f_\pi m_{\ell_i}
\left(\overline{({\ell_j}_L)^C}~{\hat p_{\nu_i}\over p^2_{\nu_i}}
{\ell_i}_R\right)K^+(p_K)\pi^-(p_\pi).
\end{equation}
Recalling that $V_{us}/V_{ud} \approx 0.2$,
$F_{K^+} \approx F_{\pi^+}$ and $f_K \approx f_\pi$,
the main contribution comes from the last matrix element with
$i=2$, because the matrix element is chirally
suppressed in the other case.
As an example, one can obtain a constraint on the 
product of $\lambda'_{2k2}\lambda'_{11k}$ and $V_{LR}$ 
from ${K^+\to \pi^- e^+\mu^+}$ decay.
In this case, we have the following numerical result
\footnote{
In  \cite{LitShr} a similar constraint was found with a different choice
of diagrams.}:
\begin{equation}  
\Gamma(K^+\to \pi^- e^+\mu^+)   =
V_{LR}^2(\lambda'_{2k2}\lambda'_{11k})^2\times
\left(\frac{100~\mbox{GeV}}{m_{\tilde d_k}}\right)^4
\times  2.5\times 10^{-28} \mbox{~GeV}
\end{equation}
or
\begin{equation}
V_{LR}(\lambda'_{2k2}\lambda'_{11k})\times
\left(\frac{100~\mbox{GeV}}{m_{\tilde d_k}}\right)^2 \le 10.
\end{equation}
It is apparent that kaon decay into a pion and a like-sign lepton pair
is too strongly suppressed to be useful at present: the corresponding
bounds on the
$\lambda'\lambda'$ product are currently of the order of 100.

\section{Conclusions}

We have discussed 
in this paper flavour-violating decays of
kaons into charged-lepton pairs in supersymmetric theories,
in both the Minimal Supersymmetric Standard Model and
$R$-violating models. In the first case,
these decays are generated by box diagrams involving
charginos and neutralinos,
and both the squark and the slepton mixings
enter in the analysis.
The process looks promising for
correlating the (s)-quark and (s)-lepton mixing
by a combined study of rare charged lepton and
kaon decays. Despite the
limits from $\mu \rightarrow e \gamma$,
$\mu$--$e$ conversion 
and $\Delta m_K$, the kaon decay branching ratios
for large $\tan\beta$ and
small soft supersymmetry-breaking terms may be accessible
to a future generation of experiments using new intense proton sources.

In the case of 
$R$-violating supersymmetry, such rare
kaon decays may occur at tree level. 
In this case, $\mu \rightarrow e \gamma$ again occurs
via one-loop diagrams, whilst $\mu$--$e$ conversion may
also occur at tree-level, but via a set of operators
different from those relevant to kaon decays.
In this framework, we studied the expected rates for
the decays $K \rightarrow \mu^\pm e^\mp$ and
$K \rightarrow \pi \mu e$, for
all two- and three-body processes. 
Using the current experimental data,
we obtained the bounds on
products of $LL\bar{E}$ and $LQ\bar{D}$ operators
summarized in (\ref{res-k1}) and (\ref{res-k2}).
We have also noted the possibility of like-sign lepton
events in the presence of non-zero $\tilde{b}_L$
--$\tilde{b}_R$ mixing, but for this to occur
at significant rate one would
need large $R$-violating couplings.

Our final conclusion is  that
lepton-flavour-violating
rare kaon decays have the potential to provide important information
on the issue of flavour physics. Any future observation would,
in addition, help distinguish between
different supersymmetric theories.

\begin{center}
{\large \bf Acknowledgements:}
We thank Gerhard Buchalla for useful discussions.
\end{center}


\begin{center}
{\large 
\bf Appendix:} {\large \bf The rate for 
$K \rightarrow \mu e$ decay via box diagrams in the MSSM}
\end{center}

The branching ratio for $K \rightarrow \mu e$ is given by:
\bea
BR (K \rightarrow \mu e) = 
\frac{2.65 \lambda^2}{2 G_F \sin^2\theta_W} 
(|K_L|^2+|K_R|^2)
\eea
where $\lambda = {(m_d+m_s)m_\mu}/{m_K^2}$,
and $K_L$, $K_R$ are given by the following
expressions:
\bea
K_L & = & K_L^c + (K_L^{n(1)}+K_L^{n(2)}) \\
K_R & = & K_R^c + (K_R^{n(1)}+K_R^{n(2)}) 
\eea
with
\begin{eqnarray}
K_L^c & = & \frac{1}{4}J_{4(A,B,X,Y)}
\left (-\frac{\lambda}{2}
 C^{R(d)}_{dAX} C^{R(d)*}_{sBX}
C^{R(\ell)}_{\mu BY} C^{R(\ell)*}_{eAY}
+ C^{R(d)}_{dAX} C^{L(d)*}_{sBX}
C^{L(\ell)}_{\mu BY} C^{R(\ell)*}_{eAY} \right ) \\
&&
-\frac{1}{4} I_{4(A,B,X,Y)} 
m_{\tilde{\chi}_A^-} m_{\tilde{\chi}_B^-}
\left (
 C^{L(d)}_{dAX} C^{R(d)*}_{sBX}
C^{L(\ell)}_{\mu BY} C^{R(\ell)*}_{eAY}
+ \lambda ~C^{L(d)}_{dAX} C^{L(d)*}_{sBX}
C^{R(\ell)}_{\mu BY} C^{R(\ell)*}_{eAY} \right )
\nonumber \\
K_R^c & = & - K_L^c |_{L \leftrightarrow R}  
\end{eqnarray}
where $m_{\tilde{\chi}_{A,B}}$ and $m_{\tilde{l}_{X,Y}}$
denote chargino and
sneutrino masses, in the chargino contribution.
Moreover,
\begin{eqnarray}
iJ_{4(A,B,X,Y)}&=&\int \frac{d^4k}{(2 \pi)^4}\frac{k^2}
{(k^2-M_{\tilde{\chi}_A}^{2})(k^2-M_{\tilde{\chi}_B}^{2})
(k^2-m_{\tilde{l}_X}^2)(k^2-m_{\tilde{l}_Y}^2)}
\\
iI_{4(A,B,X,Y)}&=&\int \frac{d^4k}{(2 \pi)^4}\frac{1}
{(k^2-M_{\tilde{\chi}_A}^{2})(k^2-M_{\tilde{\chi}_B}^{2})
(k^2-m_{\tilde{l}_X}^2)(k^2-m_{\tilde{l}_Y}^2)}.
\end{eqnarray}
whilst the mixing coefficients $C^{R,L}$ appear 
in the fermion-sfermion-chargino interaction Lagrangian:
\begin{eqnarray}
   {\cal L}_{\rm int}& = &
     \bar \ell_i (C^{R (l)}_{iAX} P_R+ C^{L(l)}_{iAX} P_L )
     \tilde \chi^-_A \tilde \nu_X
\nonumber \\
  & &+ \bar \nu_i (C^{R (\nu)}_{iAX} P_R+ C^{L(\nu)}_{iAX} P_L )
     \tilde \chi^+_A \tilde \ell_X
\nonumber \\
  & &+ \bar d_i (C^{R (d)}_{iAX} P_R+ C^{L(d)}_{iAX} P_L )
     \tilde \chi^-_A \tilde u_X
\nonumber \\
  & &+ \bar u_i (C^{R (u)}_{iAX} P_R+ C^{L(u)}_{iAX} P_L )
     \tilde \chi^+_A \tilde d_X
   +h.c.
\end{eqnarray}
and their explicit expressions are
given, for instance, in \cite{HMTY}.

The fermion-sfermion-neutralino interaction Lagrangian 
is similarly written as
\begin{equation}
  {\cal L}_{\rm int}
=  \bar f_i (N^{R(f)}_{iAX} P_R +N^{L(f)}_{iAX} P_L)
   \tilde \chi^0_A \tilde f_X
\end{equation}
where $f$ stands for $l,\nu,d$ and $u$.
The neutralino box contributions
corresponding to the permutations of the $\mu$ and $e$ in the external
lines, $K_L^n(2)$ and $K_R^n(2)$, are then found to be
\begin{eqnarray}
K_L^{n(1)} & = & \frac{1}{4}J_{4(A,B,X,Y)}
\left (-\frac{\lambda}{2}
 N^{R(d)}_{dAX} N^{R(d)*}_{sBX}
N^{R(\ell)}_{\mu BY} N^{R(\ell)*}_{eAY}
+ N^{R(d)}_{dAX} N^{L(d)*}_{sBX}
N^{L(\ell)}_{\mu BY} N^{R(\ell)*}_{eAY} \right ) \\
&&
-\frac{1}{4} I_{4(A,B,X,Y)} 
m_{\tilde{\chi}_A^0} m_{\tilde{\chi}_B^0}
\left (
 N^{L(d)}_{dAX} N^{R(d)*}_{sBX}
N^{L(\ell)}_{\mu BY} N^{R(\ell)*}_{eAY}
+ \lambda N^{L(d)}_{dAX} N^{L(d)*}_{sBX}
N^{R(\ell)}_{\mu BY} N^{R(\ell)*}_{eAY} \right )
\nonumber \\
K_R^{n(1)} & = & - K_L^{n(1)} |_{L \leftrightarrow R}  
\end{eqnarray}
and 
\begin{eqnarray}
K_L^{n(2)} & = & \frac{1}{4}J_{4(A,B,X,Y)}
\left (\frac{\lambda}{2}
 N^{R(d)}_{dAX} N^{R(d)*}_{sBX}
N^{L(\ell)}_{\mu BY} N^{L(\ell)*}_{eAY}
- N^{R(d)}_{dAX} N^{L(d)*}_{sBX}
N^{L(\ell)}_{\mu BY} N^{R(\ell)*}_{eAY} \right ) \\
&&
+ \frac{1}{4} I_{4(A,B,X,Y)} 
m_{\tilde{\chi}_A^0} m_{\tilde{\chi}_B^0}
\left (
 N^{L(d)}_{dAX} N^{R(d)*}_{sBX}
N^{L(\ell)}_{\mu BY} N^{R(\ell)*}_{eAY}
+ \lambda N^{L(d)}_{dAX} N^{L(d)*}_{sBX}
N^{L(\ell)}_{\mu BY} N^{L(\ell)*}_{eAY} \right )
\nonumber \\
K_R^{n(2)} & = & - K_L^{n(2)} |_{L \leftrightarrow R}  .
\end{eqnarray}

\end{document}